\documentclass[useAMS,usenatbib,aps,prd,twocolumn]{revtex4-1}
\usepackage{times,graphicx,amsmath,amsfonts,amssymb,aas_macros,epstopdf,ulem}
\usepackage{epsfig}
\usepackage[usenames,dvipsnames]{xcolor}
\usepackage{url}
\usepackage{subfigure}
\usepackage[colorlinks=true, urlcolor=black]{hyperref}

\def\ie{{\frenchspacing\it i.e.}}
\def\eg{{\frenchspacing\it e.g.}}

\def\be{\begin{equation}}
\def\ee{\end{equation}}
\def\ba{\begin{eqnarray}}
\def\ea{\end{eqnarray}} 
\def\Msun{h^{-1}{\rm M}_{\odot}}
\def\hmpc{h^{-1}\,{\rm Mpc}}

\def\hmpcc{h^{-3}\,{\rm Mpc}^3}

\def\dd{\textrm{d}}

\def\de{\delta}
\def\der{\delta_R}
\def\ln{{\rm ln}\,}
\newcommand{\ad}[1]{\langle\de^{#1}\rangle}
\newcommand{\av}[1]{\langle{#1}\rangle}

\def\frac#1#2{{\textstyle{#1\over #2}}}

\def\simlt{\stackrel{<}{{}_\sim}}
\def\simgt{\stackrel{>}{{}_\sim}}

\begin{document}
\title{Revealing modified gravity signal in matter and halo hierarchical clustering}
\author{Wojciech A.~Hellwing$^{1,2}$}
\author{Kazuya Koyama$^{1}$}
\author{Benjamin Bose$^{1}$}
\author{Gong-Bo Zhao$^{3,1}$}
\affiliation{$^{1}$Institute of Cosmology and Gravitation, University of Portsmouth, Portsmouth PO1 3FX, UK}
\affiliation{$^{2}$Janusz Gil Institute of Astronomy, University of Zielona G\'ora, ul. Szafrana 2, 65-516 Zielona G\'ora, Poland}
\affiliation{$^{3}$National Astronomy Observatories, Chinese Academy of Science, Beijing, 100012, P.R.China}
\date{\today}

\begin{abstract}
We use a set of N-body simulations employing a modified gravity (MG) model with Vainshtein screening to study matter 
and halo hierarchical clustering. As test-case scenarios we consider two normal branch Dvali-Gabadadze-Porrati (nDGP) 
gravity models with mild and strong growth rate enhancement. We study higher-order correlation functions $\xi_n(R)$ up 
to $n=9$ and associated hierarchical amplitudes $S_n(R)\equiv\xi_n(R)/\sigma(R)^{2n-2}$. We find that the matter PDFs 
are strongly affected by the fifth-force on scales up to $50h^{-1}$Mpc, and the deviations from GR are maximised at $z=0$. 
For reduced cumulants $S_n$, we find that at small scales $R\leq10h^{-1}$Mpc the MG is characterised by lower values, 
with the deviation growing from $7\%$ in the reduced skewness up to even $40\%$ in $S_5$. To study the halo clustering 
we use a simple abundance matching and divide haloes into thee fixed number density samples. The halo two-point functions 
are weakly affected, with a relative boost of the order of a few percent appearing only at the smallest pair separations 
($r\leq 5h^{-1}$Mpc). In contrast, we find a strong MG signal in $S_n(R)$'s, which are enhanced compared to GR. 
The strong model exhibits a $>3\sigma$ level signal at various scales for all halo samples and in all cumulants. 
In this context, we find that the reduced kurtosis to be an especially promising cosmological probe of MG. Even the mild
nDGP model leaves a $3\sigma$ imprint at small scales $R\leq3h^{-1}$Mpc, while the stronger model deviates from a GR-signature 
at nearly all scales with a significance of $>5\sigma$. Since the signal is persistent in all halo samples and over 
a range of scales, we advocate that the reduced kurtosis estimated from galaxy catalogues can potentially constitute 
a strong MG-model discriminatory as well as GR self-consistency test.
\end{abstract}


\maketitle
\section{Introduction}
\label{sec:intro}

The standard model of cosmology - Lambda Cold Dark Matter (LCDM) - is one of the biggest accomplishments
of modern physics of the last three decades. This model describes how the Universe cooled down and expanded
from the initial fireball of the Big Bang and formed the large-scale structure (LSS) observed presently in
wide and deep galaxy spectroscopic surveys. Astonishingly, LCDM, being a very simple model characterised by only six
free parameters, passes a tremendous number of robust observational tests. It explains the features and correlations
observed in the Cosmic Microwave Background (CMB) \citep[\eg][]{WMAP9,Planck1}, the primordial nucleosynthesis and light element abundance 
\citep{Yang1984,Walker1991}, 
the growth of tiny primordial density perturbations into LSS \citep{Percival2001,Tegmark2004,Beutler2017} and 
the late-time observed accelerated expansion \citep{acceleration1,acceleration2,Percival2010,Weinberg2013}. 
However, this spectacular success comes with a high price, since LCDM is mostly phenomenological in its nature.
This is because in the model, the main contributors to the cosmic energy budget are dark matter (DM) and dark energy (DE).
The physical nature of both of these constituents is far from being understood and assessed.

While the observational evidence for dark matter's existence is overwhelming, with the general consensus being that
the last piece of this puzzle is still missing due to difficulties related to hunting for the DM particle 
in Earth-based laboratories like CERN and Fermilab \citep{Ellis2000,Ahmed2009,Klasen2015}, the conceptual and theoretical problems related to DE 
are of a much more profound nature. One of the core assumptions of LCDM is that General Relativity (GR) is 
an adequate description of gravity on all cosmic scales, from the size of the Solar System ($\sim$AU) to the scale 
of particle horizon ($\sim$ Gpcs). Within this picture the only possibility to accommodate the late-time accelerated 
expansion of the Universe is
via a perfect fluid of an exotic equation of state parameter $w=-1$, dubbed Dark Energy. The simplest and most natural
candidate for DE is Einstein's cosmological constant (CC) -- $\Lambda$ -- reflecting here the zero-point energy 
associated with quantum vacuum fluctuations. Alas, identification of $\Lambda$ with DE is spoiled by the gargantuan
discrepancy between the observed tiny value of CC and the theoretical prediction from quantum field theory. 
This precipice has at least 50 orders of magnitude \citep{Carroll2001}. One of the commonly adopted solutions here consists of an arbitrary 
assumption that by some symmetry of nature the true Einstein's constant is set to exactly zero, and the observed 
accelerated expansion is due to an exotic scalar-field or other similar phenomena dominating the cosmic energy budget
at late time \citep{Copeland2006}. The main problem with this approach is that most theories experience
only weak coupling of the scalar field to matter, predicting growth of structure that is the same as in GR. This
makes these models very hard to falsify.

Another way to approach the DE problem is rooted in the observation that GR has been rigorously tested only on small
scales of the order of hundreds Astronomical Units at most \citep{Bertotti2003,Will2014,Baker2015,Brax2014,Raveri2015,Berti2015,Abbott2016}. 
Thus, using GR to describe intergalactic, cosmological
and ultimately horizon scales, is an extrapolation by a spectacular 15 orders of magnitude. Therefore, one could argue for 
a different interpretation of the late-time Universe's acceleration, which would not be due to a mysterious DE, but
could be a manifestation of the breakdown of GR on cosmological scales. Such scenarios have received large attention
in the literature over the past two decades and are commonly described as Modified Gravity (MG) theories. Here, viable theories
are usually built as infra-red modifications to GR that can fuel the low redshift acceleration of the space-time via virtue 
of modifications to the Friedman equations stemming from an altered Einstein-Hilbert action \cite{cddett2005,cfps2011}. Most of such non-trivial modifications
of GR exhibit additional degrees of freedom, propagation of which will locally manifests as an additional fifth-force acting
on test matter particles. Such a fifth force can be usually described in terms of a local effective Newtonian gravitational constant
($G_{eff}$). Notwithstanding, as mentioned earlier, we have stringent precise tests of GR on small scales, thus any prospective
MG theory needs to employ a theoretical mechanism that would allow it to recover standard GR-behaviour on Solar System scales
and around dense bodies (like black holes and neutron stars). Such an appliance is called a {\it screening mechanism} in MG parlance.
While the screening mechanism allows a given theory to pass the small-scale and strong field regime observational tests,
its introduction comes with a high price, as theories with screening exhibit much stronger non-linear behaviour \citep{bdlw2012}.

The standard cosmological model, as any other successful physical theory, is placed under continuous scrutiny.
The fact that GR has not been rigorously tested on cosmological scales, put together with the fact that on a theoretical side 
we are provided with a plethora of interesting modified theories, indicates that we are in an urgent need of precise observational
tests of gravity applied to intergalactic and cosmological scales. This worthy goal was set as the aim of many currently undergoing
and approaching observational endeavours, such as Euclid and DESI \citep{Laureijs2011,Levi2013,DESI2016} to name a few. These programmes aspire to chart
the Universe's large scale structure on vast spatial and time scales. The hope is that by analysing properties of LSS one can
measure the growth rate of structure at different epochs, which when determined with enough accuracy would constitute a strong
null-test for GR on cosmic scales \cite{Jain2010,Bean2010,Berti2015}. The validity of this test relies on the observation that within the standard paradigm
the observed LSS structures arose from tiny primordial density perturbations via a mechanism called {\it the gravitational instability}.
This mechanism explains how due to self-gravity, in an expanding background, small initial overdense regions
($\rho(\vec{x})>\langle\rho\rangle$) collapse into bound, and finally virialised, structures that host luminous 
galaxies today. Here, the theory aspires to describe the growth on tiny irregularities in the course of the cosmic expansion history
from initial density perturbations of the order of $\sim10^{-5}$, as observed in the CMB \citep{Kogut1993}, to present day DM haloes
characterised by central densities of $\simgt 10^6$ \citep[see \eg][]{Frenk1985,Makino1998,FrenkWhite2012}.
Thus, to connect the theoretical predictions with observations we need an accurate description of the growth 
of structures spanning more then 10 orders of magnitude in density.

On sufficiently large scales (\ie{} $\simgt 100\hmpc$) or interdependently at sufficiently early times ($z\simgt1$) 
the growth of structures can be described by linear and weakly non-linear perturbation theory (PT). This picture has been shown
to be accurate and has been tested rigorously in the past, both for GR and non-GR theories \citep[see \eg][]{BCGS_book}. However, the bulk amount 
of cosmological observations concerns the regime where the cosmic structure and its dynamics are deep in the non-linear regime.
The only way to probe and study the non-linear regime of gravitational instability is to use sophisticated and expensive
computer simulations. Use of simulations to study the growth of LSS has become a standard practice over the past three decades
and is now considered a mature field \citep{Davis1985,Springel2006}. 

Indeed, computer cosmological simulations have undeniably become a powerful tool of modern cosmology, but they have also introduced 
a major obstacle that makes their use for model testing difficult. Namely, for each set of initial conditions described by some
chosen values of cosmological parameters and for each specific GR/MG model, one needs to run a separate dedicated computer simulation. 
In addition, due to intrinsic non-linearity of the screening mechanism employed by MG,  dedicated numerical codes are also needed 
for running non-standard gravity simulations. Covering the non-linear regime of MG structure formation  with sufficient resolution
is also much more computationally expensive than standard GR simulations. These difficulties made the study of MG 
theories very non-trivial and challenging. Nonetheless, we need to stress out that it is absolutely necessary to use dedicated
MG simulations for rendering theoretical predictions for growth rate of structures. This is because both the complicated nature
of galaxy formation physics and the non-linear character of screening mechanisms introduce various degeneracies and biases \citep{mog_gadget}.
In order to formulate self-consistent predictions for MG signatures in an observable one needs to assess, understand and disentangle
various systematic effects from the ones that are purely a result of altered dynamics stemming from additional MG degrees of freedom \citep{Bull2016}.

The situation is not as hopeless as one might expect when considering the very rich phenomenology of many MG theories.
Many different screening mechanisms can be categorised as either being screened by gravitational potential or density. 
The first is a broad category, where the screening suppresses the fifth force by either making locally the scalar field very 
massive ({\it the chameleon mechanism})\cite{kw2004}, forcing a small value of scalar field (the {\it symmetron } fields)\cite{dlmw2012}, 
or suppressing the strength of coupling of the scalar field ({\it the dilaton fields}) to matter \cite{bbdls2011} in high density regions. 
The flagship example of the latter screening category is the {\it Vainsthein mechanism}\cite{Vainshtein:1972sx}, where due to higher-order 
derivative interactions in the vicinity of massive objects the scalar field fluctuations attain significant kinetic terms and thus decouple 
from the matter fields.
In the past decade the non-linear gravitational clustering in MG theories has received much attention 
\citep[see \eg reviews in][]{cfps2011,Koyama2016,Bull2016}.
With the chameleon screening employed by the scalar-tensor theory of $f(R)$ being by far the most thoroughly inquired theory \cite{hs2007,Sotiriou:2008rp}.
It has been shown that this class of theories exhibit very interesting behaviour at late times predicting strong observational 
features in anisotropic galaxy
redshift space clustering \cite{jblkz2012}, matter and velocity power spectra \cite{LiHell2013}, halo and galaxy dynamics \cite{Hellwing2014PhRvL},
real and redshift space halo clustering \cite{halosvoids_fr,Arnalte-Mur2017} and higher-order clustering statistics \cite{Hellwing2013}. 
On the other hand the gravitational instability mechanism with Vainshtein 
screening was probed to a much lower extent, as most of the studies concerned either only two-point matter clustering statistics 
\cite{schmidt2009,Li2013,Barreira2013}, 
some basic morphological LSS features \cite{Falck2014, Falck2015}, screening of dark matter halos \cite{Falck2015} or simple halo dynamics \cite{Hellwing2014PhRvL}, and further those studies 
were mostly based on simulations 
with both limited resolution and volume. The results presented in this paper are aimed to amend this situation and provide comprehensive 
insights into nature of gravitational instability within Vainshtein mechanism type of screened theories.

As our test-bed we chose to take the so-called {\it normal branch of Dvali-Gabadadze-Poratti} model (nDGP), a higher-dimensional gravity
brane-world model that employs Vainshtein mechanism \cite{DGP2000}. The choice of nDGP as our test case for Vainshtein is motivated by the fact
that this model, in contrast to so-called self-accelerating branch (sDGP) \cite{sDGP} or other models 
like covariant Galileons \cite{Nicolis:2008in} that employ Vainsthein,
is fully consistent with LCDM's expansion history as precisely determined by modern observations. However, we ought to mention 
that there is a major drawback here related to nDGP. In order achieve its compatibility with the data, it still requires some
small amount of DE \cite{schmidt2009}. Hence, its attractiveness as an alternative explanation of accelerated expansion is largely diminished.  
This being said, we want to stress that our main intention here is not to present and test a new fundamental theory of gravity
as an alternative explanation to cosmic acceleration, but rather to study phenomenology of a large-class of models by assessing 
the impact on cosmic structure formation of a fifth-force moderated by Vainshtein screening.

In addition to the gravitational instability, the second conventional assumption of the standard model for the formation
of structures is that the primordial density fluctuations were described by Gaussian random field statistics.
The structure formation theory is bound by providing an appropriate description of the initial power spectrum of the density fluctuations.
Now, by taking the main ingredient to be cold DM, one obtains a hierarchical model of structure formation, 
where the clustering and gravitational collapse proceeds from small to large scales. For power law perturbation spectra, $P\propto k^{n_{\rm s}}$, 
this is always true, provided $n_{\rm s}>-3$. In the nDGP gravity we assume that all elements of the structure formation model are the same as in LCDM, 
except for the non-linear modifications to gravity surfacing through the Vainshtein screened fifth force. As the enhanced fifth-force dynamics
become important at different cosmic scales and epochs, it will lead to departures from the standard well established and tested hierarchical clustering
paradigm of GR. This changes in turn should be imprinted in hierarchical clustering statistics of matter and DM haloes.
For the case of MG models with different screening mechanisms it was shown that their modified dynamics in most cases
leave strong imprints on the matter clustering hierarchy, especially in the higher order moments \cite{Hellwing2010,Hellwing2013}. 
Since the hierarchical clustering as a main prediction of the gravitational instability scenario was so thoroughly tested in 
the case of GR and some other MG models, it is now imperative to conduct such studies also for the Vainshtein class of fifth-force cosmologies. 
This defines the main goal of the analysis presented in this paper.

This paper is organised as follows: in \S\ref{sec:nDGP_desc} we provide a general description of the physical properties of the nDGP model,
\S\ref{sec:simulation} covers the details of N-body simulations used in this study. In Sec. \S\ref{sec:hierarchical_theory} we present
the main features of hierarchical clustering theory, this is followed by the main results of our analysis presented in \S\ref{sec:results}.
Finally, we conclude in \S\ref{sec:conclusions}.

\section{nDGP gravity model and simulations }
\label{sec:nDGP_desc}

\subsection{Model}

We consider the normal branch Dvali Gabadadze Porrati (nDGP) braneworld model \cite{DGP2000} that has the same expansion 
history as LCDM. Under the quasi-static perturbations, the Poisson equation is given by~\cite{Koyama2007}
\ba
	\nabla^2 \Psi & =& 4 \pi G a^2 \rho \delta + \frac{1}{2} \nabla^2 \varphi, \nonumber\\
	\nabla^2 \varphi & + & \frac{r_c^2}{3 \mathcal{B}(a) a^2}
	[(\nabla^2 \varphi)^2 -(\nabla_i \nabla_j \varphi)(\nabla^i \nabla^j \varphi)]
	= \frac{8 \pi G a^2}{3 \mathcal{B}(a)} \rho \delta, \nonumber\\
	\label{eq:phievo1}
\ea
where $\Psi$ is the Newtonian potential, $G$ is the Newton constant, $a$ is the scale factor and $\varphi$ is an additional scalar field in the model.  
The function $\mathcal{B}(a)$ is given by
\begin{equation}
	\mathcal{B}(a) = 1 +  2 H r_c \left(1+ \frac{\dot{H}}{3 H^2} \right),
	\label{eq:beta}
\end{equation}
where $r_c$ is the cross-over scale above which gravity becomes five-dimensional. Now by defining $\Omega_{r_c} \equiv1/ (2 r_c H_0)^2$ 
we obtain a single parameter that defines any given nDGP model. Here, $H_0$ is the present-day value of the Hubble parameter and we 
have followed a usual convention with $c=1$. 

On linear scales, where we can ignore the non-linearities of the scalar field, the fifth-force enhances the Newtonian gravity.
This can be quantified by means of an effective Newton constant, which will be given by $G_{eff} = G \{1+ 1/[3 \mathcal{B}(a)]\}$.
Note that $\mathcal{B}(a)$ is positive and decreasing in time, so 
the growth of structure formation is always enhanced in this model and the enhancement becomes larger at late times. For 
a larger $\Omega_{r_c}$, $\mathcal{B}(a)$ is smaller so the enhancement of gravity is stronger. On small scales, the non-linearity 
of the scalar fields by virtues of the Vainshtein screening suppresses the coupling between the scalar field and matter.
Thus the effective gravity approaches the GR case, $G_{eff}\rightarrow G$. 
This class of models experience rich phenomenology, as the Vainshtein mechanism is intrinsically non-linear.

\subsection{Simulations}
\label{sec:simulation}
\begin{table}[h!]
\caption{\label{tab:halo_samples} The number densities and the corresponding cut-off halo masses
for our halo sample at $z=0$.}
\medskip
\begin{tabular}{cccc}
\hline
\hline
number density & GR $M_{min}$ & nDGPa $M_{min}$ & nDGPb $M_{min}$\\
$\hmpcc$ & $\Msun$ & $\Msun$ & $\Msun$ \\
\hline
\hline
$1.4\times10^{-3}$ & $1.6\times10^{12}$ & $1.6\times10^{12}$ & $1.72\times10^{12}$\\
\hline 
$9\times10^{-4}$ & $1.93\times10^{12}$ & $1.94\times10^{12}$ & $1.96\times10^{12}$\\
\hline
$1\times10^{-4}$ & $2.54\times10^{13}$ & $2.57\times10^{13}$ & $2.7\times10^{13}$\\
\hline
\hline
\end{tabular}
\end{table}

In our analysis we will use a set of N-body simulations run for LCDM and two nDGP models.
These simulations were conducted using the AMR code ECOSMOG~\citep{ECOSMOG}. 
The background cosmology is taken from WMAP9~\citep{WMAP9}: $\Omega_m = 0.281$, $h=0.697$, and $n_s=0.971$. 
The box length is $1024$Mpc/h with $1024^3$ dark matter particles used and a starting redshift of $49$.
The initial conditions were generated using {\tt MPGrafic}\footnote{Available at \url{http://www2.iap.fr/users/pichon/mpgrafic.html}}.
Using $z_{ini}=49$ assures that the system will be evolved for time long enough in order to wipe-out any
transients effects that are affecting higher moments of initial particle distribution displaced 
by Lagrangian methods \cite{Scoccimarro1998,transients1,transients2}
This design sets the resulting mass resolution at $m_p\cong7.8\times10^{10}M_{\odot}h^{-1}$ and the Nyquist fluid approximation
limit of $k_{Nyq}\cong\pi$ h/Mpc.The most refined AMR grid were at the level 16, setting a maximal force 
resolution at $\epsilon=1024/2^{16}=0.015$Mpc/h.
The LCDM run will constitute our fiducial GR-reference point, in addition we simulate two nDGP models implementing
two values of the cross-over scale parameter: $\Omega_{rc} = 0.0124, 0.438$. The first model (nDGPa) 
is characterised by only a mildly enhances growth of structure history and we will treat is as a borderline case. 
The latter model (nDGPb) with the large value of $\Omega_{rc}$ parameter should experience 
sizeably larger differences from the GR case, fostering a more realistic detection prognosis.

We evolve the dark matter density and velocity field from the initial redshift to the present epoch, selecting and saving
for the analysis snapshots taken at three specific epochs: $z=0,0.5$ and $1$. To identify DM haloes in our snapshots 
we resort to the {\tt ROCKSTAR} FOF phase-space halo finder \cite{Behroozi2013}. We keep all the haloes and subhaloes with 20 or more
particles for further analysis. As a main proxy for halo mass we settle down for a commonly used virial mass
$M_{200}\equiv 4/3\pi R_{200}^3 200\times\rho_{c}$. Where $R_{200}$ is a boundary radius at which the spherically
averaged matter density enclosed inside is equal to 200 times the Universe critical density $\rho_{c}$. 
At $z=0$ for all three models we find a very similar number of $'\sim1.5\times10^6$ distinct haloes and subhaloes,
with a very similar satellite fractions of $\sim6\%$ for GR and mild nDGP models and $7\%$ for the strong nDGP model.

Once we constructed our halo catalogues we split them into samples of a fixed number density. This allows us to
approximate in a very simplistic way mock galaxy samples in the spirit of abundance matching \cite{Vale2004}. For our richest
and most complete sample we pick centrals+satellite with a number density of $\av{n}=1.4\times10^{-3}(\hmpc)^{-3}$.
We also consider sparser centrals-only samples with an effective number densities of $9$ and $1\times10^{-4}(\hmpc)^{-3}$.
The specific details like the minimum halo mass cut-off for each sample are given in Tab.~\ref{tab:halo_samples}.
\begin{figure}
 \includegraphics[angle=0,width=0.48\textwidth]{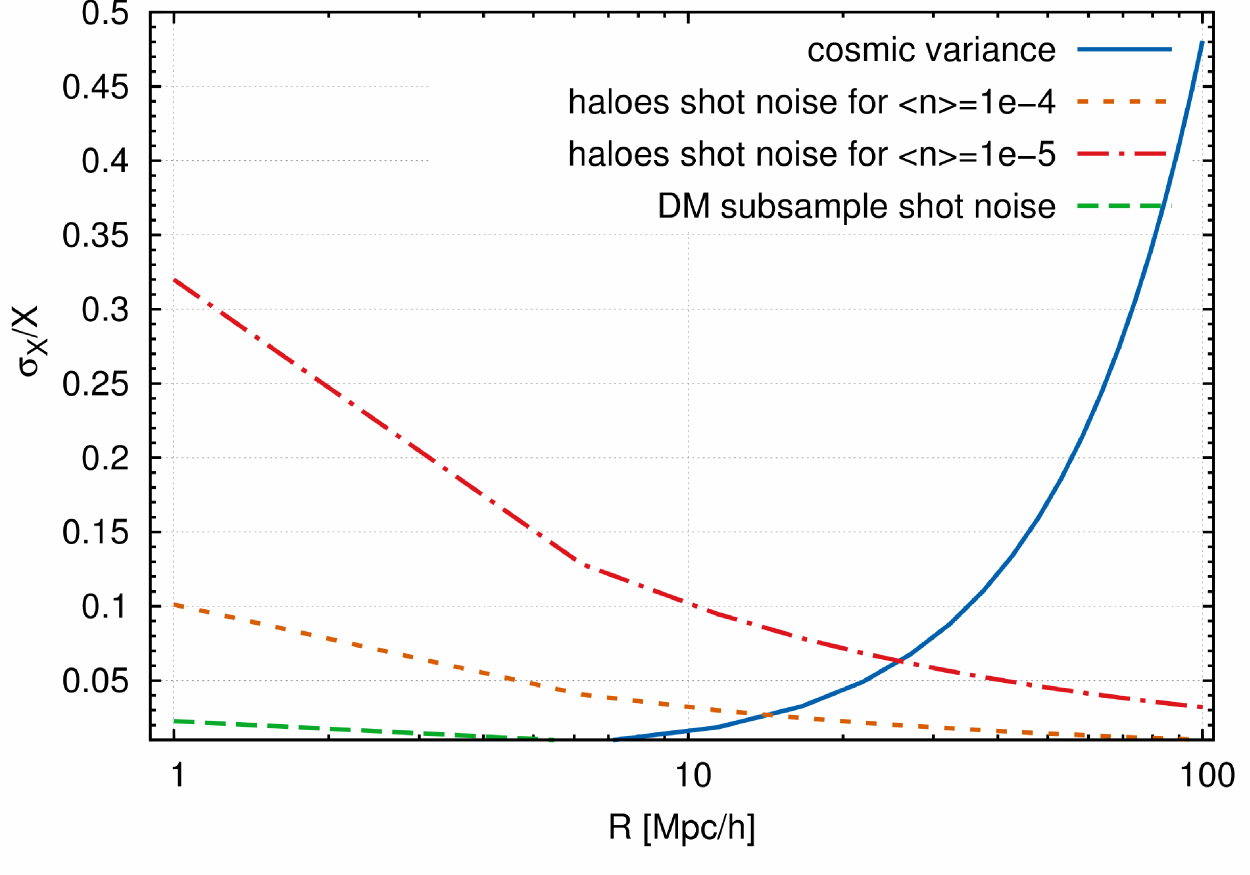}
 \caption{The contributions of the cosmic variance and the shot noise to the correlation functions 
 estimators we use in this work.}
\label{fig:errors}
\end{figure}

\section{Hierarchical clustering}
\label{sec:hierarchical_theory}

Here, our main focus will be on the matter and halo density fields, which we will describe in terms of the density contrast,
a quantity that measures local departure from a background uniform density. Thus we define
\be
\label{eqn:de_contrast}
\rho(\vec{x},t) = \langle\rho(t)\rangle\left[1+\delta(\vec{x},t)\right]\,,
\ee
where $\langle\rho(t)\rangle$ is the average background density of given tracers (matter or haloes),
and $\delta(\vec{x},t)$ (the density contrast) characterises local deviations from the background.
Now, the full statistical information about the density field and all its correlation's properties is encoded
in the density probability distribution function $p(\delta)$. 

Cosmologies employing cold dark matter spectra for initial Gaussian fluctuations
exhibit so-called {\it hierarchical clustering}. Here, the first structures to emerge
from expanding smooth background are tiny haloes corresponding to the smallest density peaks with sizes just above the dark matter
streaming scale \cite{1980Peebles,Davis1985}. As the Universe expands, larger and larger density perturbations reach the turn-around
radius and start to collapse to form bigger structures. Simultaneously, some small haloes that formed earlier
become satellites of bigger and younger structures and eventually sink and merge into them. Thus structure formation proceeds from
small to large scales. In this picture the gravitational interactions  during the evolution drive away the density probability 
distribution function from its initial Gaussian characteristics. This is reflected by the growth of higher-order
moments of the density field, which measure departure from Gaussianity \cite{Juszkiewicz1993,skew_cur_pt,grav_inst_pt}.

The first and most basic statistics that characterise clustering is the two-point correlation function: $\xi(r)$.
This is defined as the excess probability (with respect to a Poisson process) of finding 
two matter particles (or haloes) contained in two volume elements d$V_1$ and d$V_2$ at a distance $r$ (see \eg \citep{1980Peebles}):
\begin{equation}
\label{eqn:2pcf-definition}
\textrm{d}P_{12}(r) \equiv \bar{n}^2[1+\xi(r)]\textrm{d}V_1\textrm{d}V_2\,,
\end{equation}
where $\bar{n}$ is the mean matter (halo) number density. In general $\xi(r)$ characterises the strength
of matter (halo) clustering across cosmic scales and epochs. However, as mentioned above the gravitationally induced
evolution gives rise to significant departures of matter and halo density distribution functions from a normal one .
For a non-Gaussian $p(\de)$, the knowledge of only the second moment is no longer enough to fully characterise the field,
as Wick's theorem no longer holds.

In this context, the so-called reduced moments or cumulants of the distribution function $p(\delta)$ are especially useful.
The $n$-th cumulant of the distribution function $p(\de)$ is defined by recursive relation to the $n$-th
moments. This relation can be expressed by the cumulant generating function \citep[eg.][]{Lokas_kurt}
\be
\label{eqn:gen_func}
\ad{n}_{\rm c}\equiv M_n = {\partial^n \ln\av{e^{t\de}}\over \partial t^n}\bigg|_{t=0}\,.
\ee
The cumulants now can be expressed in terms of the central moments, specifically, for the first five cumulants we have \citep{ber1994,GaztanagaAPM94}
\ba
\label{eqn:cumulants}
\ad{}_{\rm c} &=& 0,\,\,\textrm{(the mean)}\nonumber\\
\ad{2}_{\rm c} &=& \ad{2}\equiv\sigma^2,\,\,\textrm{(the variance)}\nonumber\\
\ad{3}_{\rm c} &=& \ad{3},\,\,\textrm{(the skewness)}\nonumber\\
\ad{4}_{\rm c} &=& \ad{4} - 3\ad{2}_{\rm c}^2,\,\,\textrm{(the kurtosis)}\nonumber\\
\ad{5}_{\rm c} &=& \ad{5} - 10\ad{3}_{\rm c}\ad{2}_{\rm c}\,.
\ea
Here the $n$-th cumulant is obtained by taking the value of the $n$-th moment of the distribution $p(\de)$
and removing from it the contributions from all the decompositions of a set of $n$ points in its subsets
multiplied (for each decomposition) by the cumulants corresponding to each subset \citep{ber1994}.

For a field described by a normal distribution with a zero mean all cumulants, but the variance $\sigma^2$, vanish. In the standard
random field theory, the first two non-disappearing connected moments above variance have special meaning as they describe specific
shape departures of the distribution function from a Gaussian. The skewness describes the asymmetry of the 
distribution function and the kurtosis details the flattening of tails with respect to a Gaussian. Higher-order
moments characterise even more convoluted aberrations of the distribution function shape. 

Various studies of the higher order cumulants of the cosmic density fields have revealed that for the case of initial adiabatic Gaussian 
density perturbations described by a power-law spectrum, the gravitational instability mechanism produces a quasi-Gaussian clustering hierarchy
of connected moments \cite{ber1992}. Moreover, in the linear and weakly non-linear regime this hierarchy is preserved 
by gravitational time evolution \cite{Fry1984a,Fry1984b}. Commonly, this hierarchy is described in terms of the so-called {\it hierarchical scaling} relations:
\be
\label{eqn:hier_amplitudes}
S_n\equiv {\ad{n}_c\over\ad{2}_c^{n-1}}\,.
\ee
Here $S_n$ denote {\it reduced cumulants} (also called hierarchical amplitudes), and for unsmoothed fields these are constant. In reality however,
one always deals with density fields that are smoothed at some given scale. So what can be actually measured from observations and compared 
to predictions from N-body simulations are actually the {\it volume averaged} $n-point$ moments (correlation functions) of the underlying
smoothed density field $\de_R$. Now, if we define a smoothed field as
\be
\label{eqn:smoothed_de}
\de_R(\vec{x})=\int\de(\vec{x'})W(|\vec{x}-\vec{x'}|/R_w)\dd^3x'\,,
\ee
with $W(x/R_w)$ being a spherically symmetric smoothing window. For practical reasons it is convenient to limit our analysis to only
one window function that is easily applied to numerical simulations. 
Namely, we will deal with density fields smoothed over a ball of radius
$R=R_w$ with a window function normalised to unity 
\ba
\int W(y)\dd^3y = 1\quad \textrm{and}\\\nonumber
\int W(y)y^2\dd^3y = R_{\rm w}^2\,.
\ea
This filter function describes what is commonly known as the {\it top-hat} filtering. 
Now the volume averaged n-point connected moments can be defined as
\begin{flalign}
\label{eqn:xi_volume_av}
&\overline{\xi}_n(R)\equiv\av{\der^n}_{\rm c} =&&\\\nonumber
&= \int \dd^3 x_1\ldots\dd^3 x_n\xi(\mathbf{x_1}\ldots\mathbf{x_n})W(x_1/R_{\rm TH})\ldots W(x_n/R_{\rm TH})&&.
\end{flalign}

Classical PT for GR cosmology predicts that the hierarchical amplitudes computed for smoothed density fields should be weakly monotonic
decreasing functions of the smoothing scale $R$ \cite{BCGS_book}. This also has been also widely confirmed by 
comparisons with N-body simulations \cite{npoint_omega_cdm}.
The scale-dependence of the reduced cumulants is a function of increasingly complicated combinations of growing order derivatives 
of the matter variance taken effectively at a given smoothing scale. For example for the reduced skewness and kurtosis this was shown to be \citep{ber1994}
\ba
\label{eqn:s3s4_gamma}
S_3 = {34\over 7} + \gamma_1\,,\nonumber\\
S_4 = {60712\over 1323} + {62\over 3}\gamma_1 + {7\over 3}\gamma_1^2+{2\over 3}\gamma_2\,,
\ea
with $\gamma$ factors enumerating the variance derivatives
\be
\label{eqn:pt_gamma}
\gamma_n(R)\equiv {\dd^n\textrm{log}\sigma^2(R)\over \dd\textrm{log}^n R}\,.
\ee
In the Einstien-De Sitter Universe (\ie{} with $\Omega_m=1$) the skewness and kurtosis are constants, as
effectively on all scales $\gamma_n\rightarrow 0$. In LCDM however,
the $\gamma$ factors represent corrections to the reduced cumulants due to varying with scale shape of the matter power spectrum 
\citep[see \eg][]{Juszkiewicz1993,BCGS_book}.

In the nDGP model the power spectrum is modified at non-linear scales by the fifth-force dynamics \cite{schmidt2009}. Thus we can expect that this
effect should also be reflected in higher-order moments. This effect was found to be significant for other MG models like $f(R)$ \cite{Hellwing2013}
and ReBEL \cite{Hellwing2010}, where the degree of deviation from the GR prediction increased with the order of the cumulant.

We can expect that this well understood picture can get complicated, if we consider haloes rather then the matter density fields. DM haloes
are biased tracers of the underlying smooth density fields. Therefore, for the case of reduced cumulants the hierarchical scaling 
relations will not only be simple functions of the power spectrum derivatives, but also will be described by 
higher-order hierarchical biasing \cite{FG1993,npoint-halos}.
In general, we can expect that the higher-order bias can be a complicated function of scale, time and halo mass. However, as in nDGP there are
no environmental effects, the higher-order bias should take roughly the same time and scale dependence as in GR, with only the halo mass
being the major difference driver. A detailed analysis of the hierarchical biasing in nDGP (and in MG theories in general)
is well beyond the scope of this paper, as it would merit a whole dedicated study. Thus, we leave it for future work, and will not discuss further
the bias issues in the current paper.

\subsection{PT Variance and Skewness estimators}
\label{subsec:PT_formalism}
To test our numerical results we will benchmark them against the estimators available in the context of the Eulerian perturbation theory.
Our approach will be to use a perturbative formalism to calculate the variance and reduced skewness of the matter fields 
\cite{1980Peebles,Goroff:1986ep,Juszkiewicz1993}. 
Here, the additional non-linear evolution of the density contrast is modelled by also tracing the contribution from the peculiar velocity ($v_p(\vec{x})$)
divergence field:
\be
\label{eqn:expansion_scalar}
\theta(\vec{x})={\nabla\cdot v_p(\vec{x})\over aH(a)f}\,\,,
\ee
where $H(a)$ is the Hubble function and $f\equiv \dd\ln{D_+}/\dd\ln{a}$ is the logarithmic growth rate. Here, the scale factor $a$ is used
as a mean cosmic time variable and $D_+$ is the growing mode of the linear perturbation theory solution.

For the sake of brevity and simplicity it is very convenient to express all relevant quantities and work in Fourier space.
Thus, we further define 
\be
\label{eqn:delK}
\delta(\vec{k}) \, \equiv \, (2\pi)^{-2/3}\,\int \delta(\vec{x})\,e^{-i{\vec{k} \cdot \vec{x}}}\,d^3{\vec{x}}\,,
\ee
and
\be
\label{eqn:delK}
\theta(\vec{k}) \, \equiv \, (2\pi)^{-2/3}\,\int \theta(\vec{x})\,e^{-i{\vec{k} \cdot \vec{x}}}\,d^3{\vec{x}}\,,
\ee
respectively for the density contrast and the velocity divergence fields.

The classical approach is to solve the continuity and Euler equations for $\de(\vec{k})$ order by order 
(for a comprehensive review see \eg~\cite{BCGS_book}). The evolution equations are expressed below as follows:
\begin{widetext}
\ba
\label{eqn:pt_evolution1}
a \frac{\partial \delta(\vec{k})}{\partial a}+\theta(\vec{k}) = 
-\int\frac{d^3\vec{k}_1d^3\vec{k}_2}{(2\pi)^3}\delta_{\rm D}(\vec{k}-\vec{k}_1-\vec{k}_2)
\alpha(\vec{k}_1,\vec{k}_2)\,\theta(\vec{k}_1)\delta(\vec{k}_2),
\label{eq:Perturb1}
\ea
\ba
\label{eqn:pt_evolution2}
 a \frac{\partial \theta(\vec{k})}{\partial a}+
\left(2+\frac{a H'}{H}\right)\theta(\vec{k})
-\left(\frac{k}{a\,H}\right)^2\,\Phi(\vec{k})= 
-\frac{1}{2}\int\frac{d^3\vec{k}_1d^3\vec{k}_2}{(2\pi)^3}
\delta_{\rm D}(\vec{k}-\vec{k}_{12})
\beta(\vec{k}_1,\vec{k}_2)\,\theta(\vec{k}_1)\theta(\vec{k}_2)
\label{eq:Perturb2}
\ea
\end{widetext}
where the prime denotes a derivative w.r.t the scale factor $a$, $\vec{k}_{12} = \vec{k}_1+\vec{k}_2$, $\delta_D$ 
is the Dirac delta function and the mode coupling kernels, $\alpha$  and $\beta$, are given by
\begin{eqnarray}
\label{eqn:pt_kernels}
\alpha(\vec{k}_1,\vec{k}_2)=1+\frac{\vec{k}_1\cdot\vec{k}_2}{|\vec{k}_1|^2},
\quad \beta(\vec{k}_1,\vec{k}_2)=
\frac{(\vec{k}_1\cdot\vec{k}_2)\left|\vec{k}_1+\vec{k}_2\right|^2}{|\vec{k}_1|^2|\vec{k}_2|^2}.
\label{alphabeta}
\end{eqnarray}
The linearised equations are given by setting $\alpha=\beta=0$ and expanding the Poisson term, $k^2\Phi$, to linear order 
in the density perturbations. Gravity effects the evolution of the perturbations through the Newtonian potential $\Phi$ 
and one can encode any modifications to gravity there. The general order solutions for the density contrast can then be expressed as  
\begin{align}
\delta_n(\vec{k}; a) =& \int d^3\vec{k}_1...d^3 \vec{k}_n \delta_D(\vec{k}-\vec{k}_{1...n}) \nonumber \\ 
& \times F_n(\vec{k}_1,...,\vec{k}_n; a) \delta_0(\vec{k}_1)...\delta_0(\vec{k}_n) \label{nth1} 
\end{align}
where the $n^{th}$ order kernel $F_n$ is obtained by recursively solving Eqn.(\ref{eq:Perturb1}) and Eqn.(\ref{eq:Perturb2})
at $n^{th}$ order in the density contrast and velocity divergence. $\delta_0$ is the initial density contrast 
which we  assume is Gaussian. We solve for these kernels numerically at linear and second 
order within nDGP gravity and GR. This is done using the tool described in \cite{Bose:2016qun}. 

To get the smoothed fields we simply perform a Fourier transformation on Eqn.(\ref{eqn:smoothed_de}) 
\be 
\label{eqn:smothed_de_k}
\de(R) = \int\frac{\dd^3k'}{(2\pi)^3}\de(\vec{k'})W(k'R)\,,
\ee
where $W(k'R)$ is the Fourier transform of the smoothing function. For top-hat smoothing this is simply 
\be 
\label{eqn:fourier_top_hat}
W(k'R) = \frac{3}{(k'R)^3}(\sin(k'R)-k'R\cos(k'R))
\ee
The variance of the smoothed fields is then
\be 
\label{eqn:sig2_pt_estimator}
\sigma^2(R) = \int \frac{\dd^3k}{(2\pi)^3} W(Rk)^2 F_1(\vec{k};a)^2 P_L(k)
\ee
where the initial linear power spectrum, $P_L(k)$ is defined as 
$\delta_D(\vec{k}+\vec{k}')P_L(k) = \langle \de_0(\vec{k}) \de_0(\vec{k}') \rangle$. 
As the nDGP gravity effects at very early times are negligible, for both models we can use
the LCDM initial linear power spectrum. This can be calculated for a given cosmology using a Boltzmann 
solver code such as CLASS \cite{Blas:2011rf}.

On top of the variance we also consider the third moment, expressed as reduced skewness. 
By using the properties of Gaussian random fields we have \citep{1980Peebles} $\ad{3}_c = 3 \langle (\delta^{(1)})^2 \delta^{(2)}\rangle$. 
In terms of the generalised density contrast kernels and linear power spectrum this can be expressed as
\begin{align}
\ad{3}_c =  3&\int \frac{\dd^3k_1}{(2\pi)^3} \frac{\dd^3k_2}{(2\pi)^3} P_L(k_1)P_L(k_2) F_1(k_1;a) W(Rk_1) \nonumber \\  
\times & \big[ 2W(Rk_2)W(Rk_{12})F_1(k_2;a) F_2(\vec{k_1},\vec{k_2};a) \nonumber \\ &+ W(0)W(Rk_1)F_1(k_1;a)F_2(\vec{k_2},-\vec{k_2};a) \big] \label{eq:3point}
\end{align}
Note that in GR and nDGP the last term (the $3^{rd}$ line) of Eqn.(\ref{eq:3point}) vanishes. Finally, performing some of the integrals we 
obtain the following expression for the reduced skewness 
\begin{widetext}
\ba
\label{eqn:s3_pt_estimator}
S_3 = 3\int dk_1dk_2du k_1^2 k_2^2 P(k_1)P(k_2)F_1(k_1;a) F_1(k_2;a) F_2(k_1,k_2,u;a)W(Rk_1)W(Rk_2)W(Rk_{12}) \Bigg/ \nonumber \\
\bigg[ \int dk k^2 P(k)F_1(k;a)^2W(kR)^2\bigg]^2
\ea
\end{widetext}
We will use the estimators of Eqn.(\ref{eqn:sig2_pt_estimator}) and Eqn.(\ref{eqn:s3_pt_estimator}) computed for the initial power spectrum 
sued in our N-body simulations, together with $F_1$ and $F_2$ evolution kernels expressed specifically for nDGP and GR 
(see \cite{Bose2017} for details) as our PT prediction for the skewness and variance of the matter density field.

\subsection{Numerical methods}
\label{subsec:numerical_methpds}

We will specifically consider matter and halo auto-correlation function and also volume averaged moments. For the 2-point correlation
function (2PCF) we will use a simple estimator
\ba
\label{eq:2p_estimator}
  \xi(r) = {DD(r) \over N \bar{n} v(r)} -1 \, ,
\ea
where $DD(r)$ is the number of pairs of tracers with separation in the range $[r, r+\Delta r]$, $N$ is the total 
number of tracers in the sample, $\bar{n}$ is their number density, and $v(r)$ is the volume of a spherical shell 
of radius $r$ and width $\Delta r$. As we will deal only with N-body simulation data the sample selection functions in all cases 
are complete, isotropic, and homogeneous. In addition, because we have periodic boundary conditions, there are no edge effects. 
This makes the estimations of $\xi(r)$ for matter and halo samples straightforward. We rely on a numerically efficient publicly 
available package {\tt CUTE}\footnote{Available here \url{http://members.ift.uam-csic.es/dmonge/CUTE.html}} \cite{cute}.

For the case of volume averaged moments we adopt the fast method presented in \cite{Hellwing2013}. Here the density field for a given sample
is estimated using {\it the Delaunay Tesselation Field Estimator} method (DTFE)\citep{sv2000,vs2009} implemented in the publicly available
software {\tt DTFE} provided by \citep{cv2011}. Once the density field on a regular grid with $N_g$ cells have been constructed 
we proceed to apply a range of top-hat smoothing (using a FFT method) and calculate the central moments for smoothed fields:
\ba
\label{eqn:central_moments_numerical}
\av{\der^n} = {1\over N_{\rm g}}\sum_{i}^{N_{\rm g}}\left(\der^i - \av{\der}\right)^n\,.
\ea
Then we proceed to obtain connected moments and reduced cumulant using the relations of Eqn. (\ref{eqn:cumulants})-(\ref{eqn:xi_volume_av}).

In the general case our correlation function estimators will be affected by two sources of errors : the shot noise due to sparse sampling
and classical cosmic variance. Since we have only one realisation for the initial perturbation phases, we adopt a standard Poisson estimator
for the cosmic variance error, which here is a function of the number of independent density modes present in the simulation box at a given scale.
To estimate the shot noise we use the same estimator with the number counts per bin as the sampling indicator. In the case of our DM samples
the shot noise for nearly all scales (\ie{} $R>2\hmpc$) is sub-dominant to the cosmic variance contribution. 
For haloes we find that the cosmic variance is dominant at large scales of $\simgt 10-30\hmpc$ 
(depending on the halo sample number density), while at smaller scales
the shot noise dominates. We illustrate our error budget contributions in the Fig.~\ref{fig:errors}

\section{Results}
\label{sec:results}
\begin{figure}
 \includegraphics[angle=0,width=0.48\textwidth]{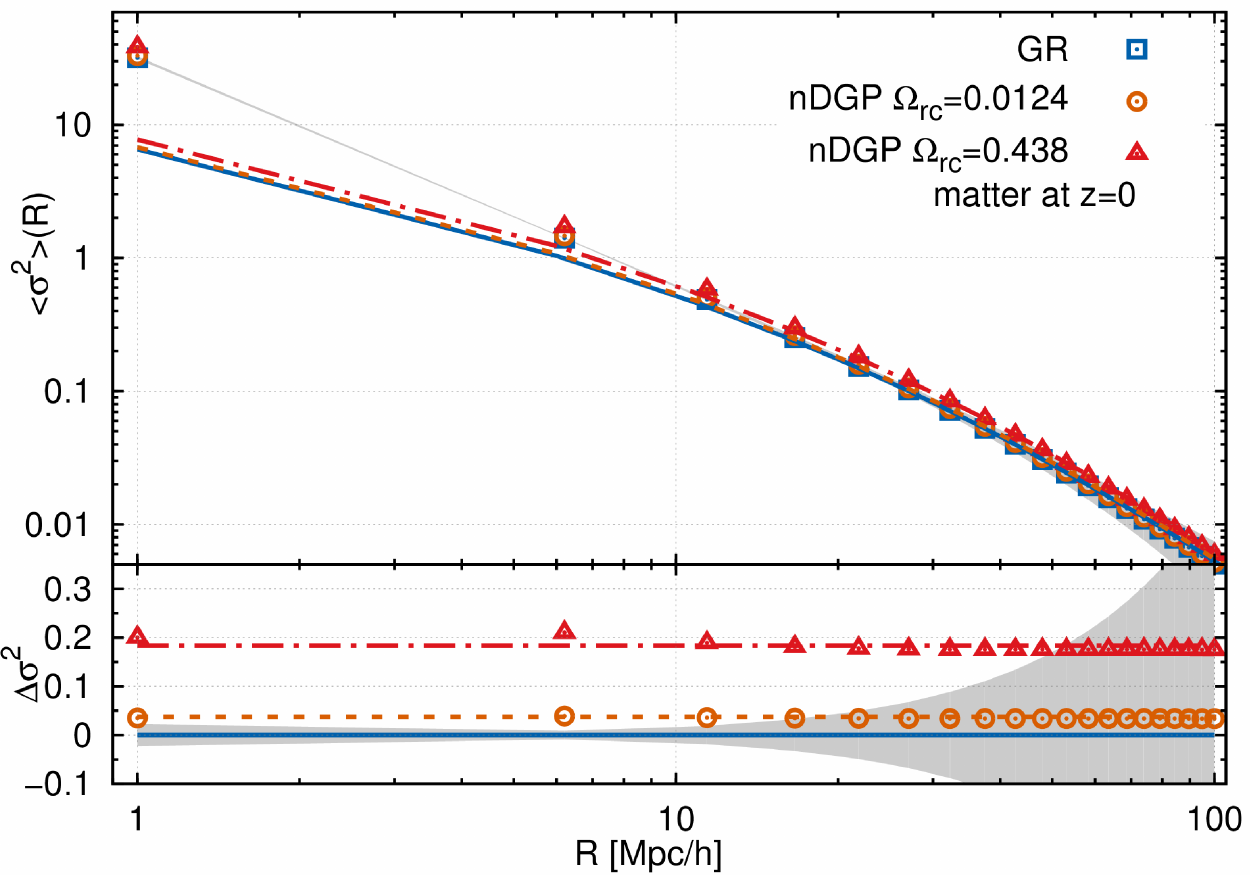}
 \caption{Comparison of matter density variance for $z=0$ estimated from N-body simulations (points) with
  the PT predictions of Eqn.(\ref{eqn:sig2_pt_estimator}). The lower panel shows the fractional difference of both 
  nDGP models taken w.r.t the GR case. The shaded region illustrate the cosmic variance error for the GR fiducial case.}
\label{fig:variance_PT}
\end{figure}
\begin{figure}
 \includegraphics[angle=0,width=0.48\textwidth]{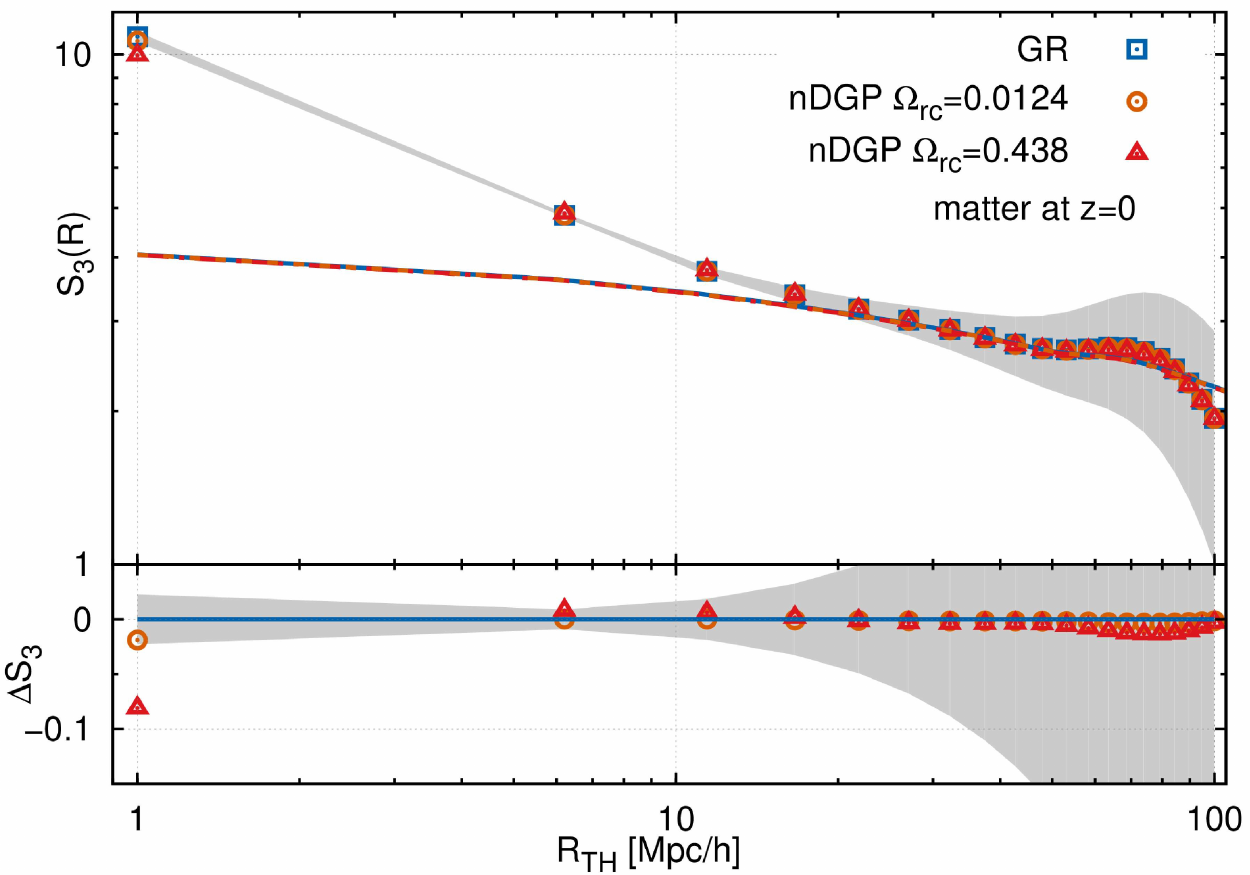}
 \caption{Same as the previous figure, but for the reduced density skewness $S_3$ compared against the estimator
 of Eqn.~(\ref{eqn:s3_pt_estimator}).}
\label{fig:skewness_PT}
\end{figure}
In the following we present the results of our analysis, where we use two-point correlations functions
alongside the higher-order reduced cumulants as well as the density distribution functions themselves,
to study the gravitational instability mechanism and the hierarchical clustering it induces for the matter and halo
density fields estimated for a range of scales at three distinctive cosmic epochs $z=0, 0.5$ and $1$.

\subsection{Dark Matter}
\label{subsec:dm}
We start by focusing on the dark matter density field. First, we want to test the implementation of our
numerical moments estimators. We do this by comparing our N-body measurements for the variance, $\sigma^2$,
and the reduced skewness, $S_3$, with the PT predictions. The results of this procedure are shown in 
Figs.~\ref{fig:variance_PT} and ~\ref{fig:skewness_PT}. Both figures illustrate a strikingly good match
of PT and N-body predictions, which agree nearly perfectly down to $R=10\hmpc$ for variance and 
down to $\sim15\hmpc$ for the reduced skewness. This is the case for all three models studied here.
Below these scales the non-linearity in the density fields become severe and PT starts to significantly 
underpredict the moments. However, the good agreement at scales that are already mildly non-linear 
reassures that our numerical estimators are accurate and unbiased.

Interestingly, we can notice that 
when we focus only on the deviations of nDGP models from the fiducial GR predictions, as quantified by
fractional differences ratios ($\Delta X= X_{nDGP}/X_{GR}-1$) shown in the lower panels, the PT predictions
remain surprisingly accurate down to the smallest scales we consider ($1\hmpc$). 
While the PT fails to capture all the effects of the non-linear gravitational evolution at small-scales
and underpredicts the absolute amplitudes of the cumulants, this failure has a universal character
for all our models. This behaviour most likely reflect the fact that both of our nDGP models exhibit only
very mild deviations from GR gravitational dynamics.

\subsubsection{1-point statistics}
\label{subsec:dm:pdfs}
\begin{figure*}
 \includegraphics[angle=0,width=0.96\textwidth]{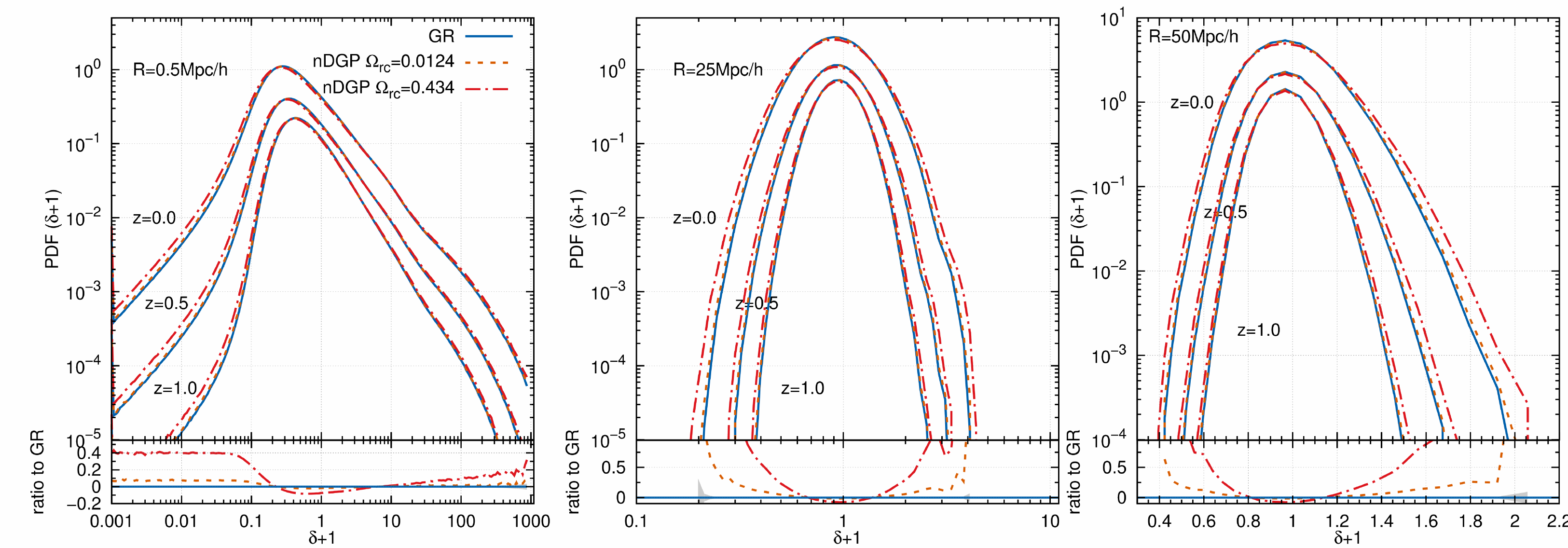}
 \caption{The distribution functions for our models computed for density fields smoothed at $R=0.5, 25$ and $50\hmpc$ (panels from left to right).
 For each panel the GR case is marked by solid blue line, nDGP ${\Omega}_{rc}=0.0124$ is dashed orange and nDGP ${\Omega}_{rc}=0.438$
 is marked by dot-dashed line. In each upper panel three groups of lines correspond to three redshifts $z=0,0.5$ and $1$, where the data for latter two
 was scaled down for clarity. For each case the smaller bottom panel illustrates the $p(\delta)$ ratio taken w.r.t the GR case only for $z=0$ case.
 Mark the change to linear scaling for the $\delta+1$ axis in the most left panel.}
\label{fig:pdf_compare}
\end{figure*}
\begin{figure}
 \includegraphics[angle=-90,width=0.48\textwidth]{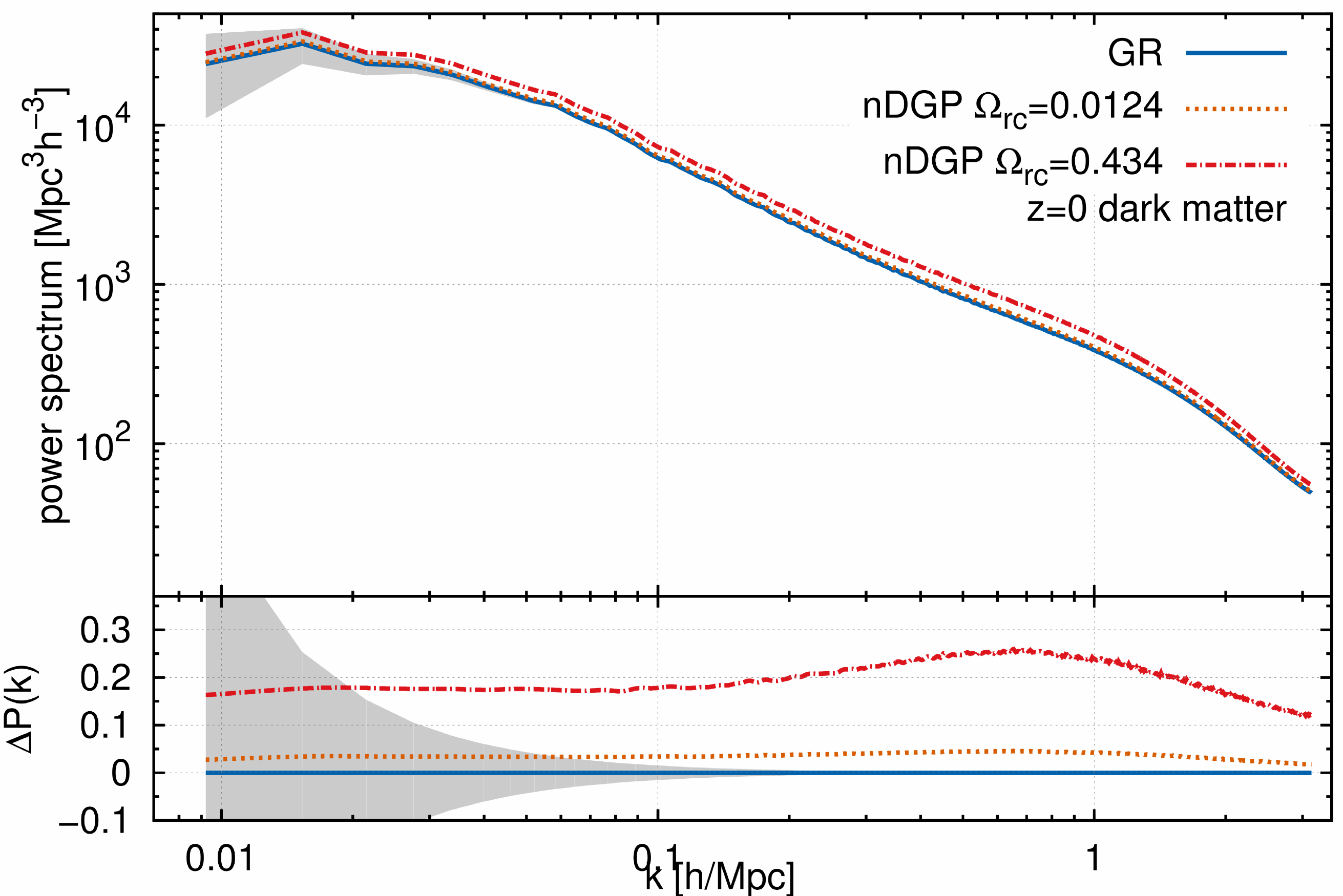}
 \caption{The matter density power spectrum computed at $z=0$ for our fiducial GR model (solid line)
 and two nDGP flavours (dotted and dashed-dotted lines). The shaded region illustrate the cosmic variance error.
 The bottom panel illustrates the fractional difference of both MG models w.r.t. the GR case.}
\label{fig:matter_pk}
\end{figure}
It is very illustrative to begin our analysis by taking a close look at 1-point statistics of the cosmic density field. This is readily characterised
by 1D density probability distribution functions (PDFs), $p(\delta+1)/(\Delta\delta)$. For a primordial density perturbation field described by adiabatic 
fluctuations these functions are Gaussian on all scales. For evolved fields the distribution functions develop exponential tails and at sufficiently 
large scales can be described by log-normal distributions \cite{Bernardeau1995}. In all cases the linear and non-linear gravitational evolution 
driving the structure formation
is encoded in the shape departures of the density $p(\delta)$  from a Gaussian. 

To get a first intuition of the differences caused by modified nDGP dynamics we show the 1D density PDF's in Fig.~\ref{fig:pdf_compare}.
Here, the three panels from left to right correspond to three different top-hat smoothing lengths applied to our reconstructed density fields: 
$R=0.5, 25$ and $50\hmpc$ respectively. 
For all scales we show the PDFs obtained at three different epochs for $z=0, 0.5$ and $1$, where the lines corresponding to $z>0$
are tiered down for clarity. First, as expected by following the previous results from PT and N-body simulations 
\cite{schmidt2009,Koyama2007}, we notice that  the difference from the fiducial GR case is maximised at the present time. 

Another interesting feature illustrated by Fig.~\ref{fig:pdf_compare} is the broadening of the PDFs towards low density ($\de<0$) values. This feature 
is present at all smoothing scales we consider. It clearly indicates that we can expect deeper density profiles of cosmic voids, which also should get larger,
as matter is more effectively evacuated towards surrounding higher density areas. While this is a very promising  feature, and potentially could leave
an observable imprint in \eg void lensing potentials, we will not study this here further, leaving this interesting opportunity for a future work.
Additionally, we can observe that at larger smoothing scales the high-density tails of the nDGP distribution functions get significantly enhanced. 
This feature reflects a simple conservation of mass, as the matter that was evacuated from the voids has to be deposited somewhere. 
This is reflected in an increased virial mass of big haloes (\ie{} cluster and supercluster scales), 
as was noted in \cite{halosvoids_fr,Hellwing2010boosting,Hellwing2013Halos}. Hence, this compensation
effect is much better appreciated at large smoothing scales, as $R_{th}=0.5\hmpc$ is way too small to encompass cluster and super-cluster density
perturbations. The general impression we can obtain by this analysis is that reduced cumulants of higher-order, like for example kurtosis,
should bear the nDGP signature at both small and large scales, while for the case of the PDF asymmetry measure, the skewness, we can expect that 
the MG-induced differences should be much more pronounced at smaller scales. We will get back to this observation later, when we discuss
the higher-order density field statistics.

\subsubsection{2-point statistics}
\label{subsec:haloes:2pt_stat}
\begin{figure*}
 \includegraphics[angle=0,width=0.48\textwidth]{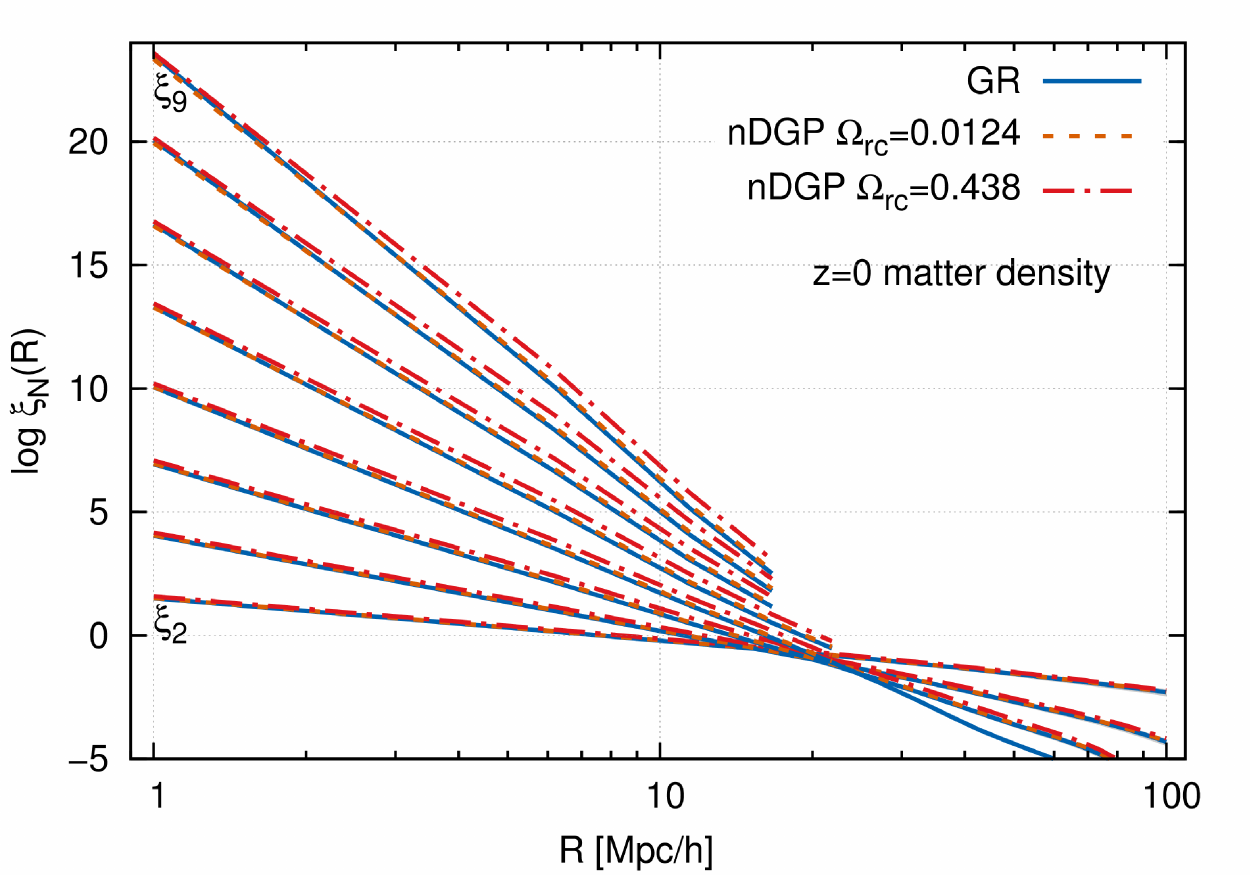}
 \includegraphics[angle=0,width=0.48\textwidth]{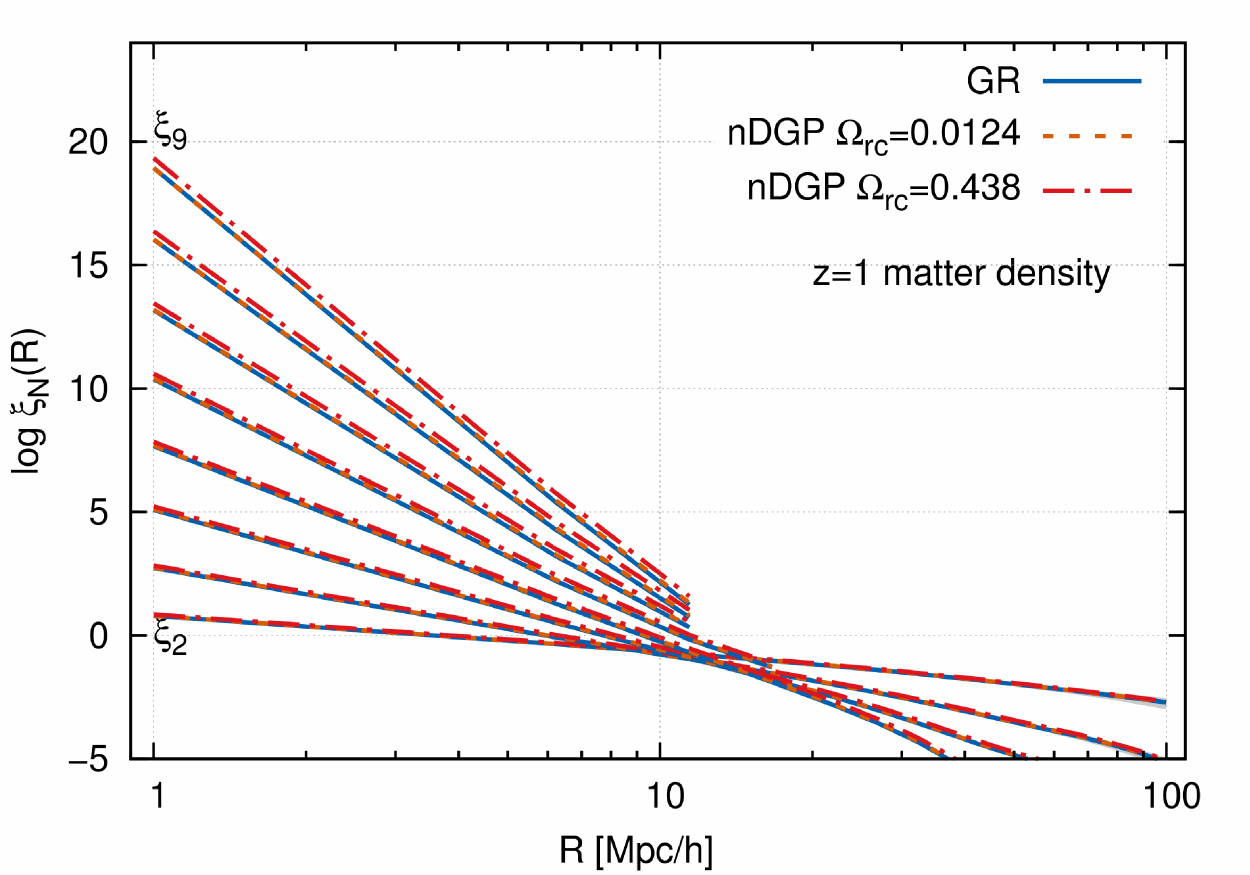}
 \caption{Hierarchy of n-point volume averaged correlation functions $\overline{\xi}_n(R)$ computed
  for smoothed dark matter density fields at $z=0$ (left panel) and $z=1$ (right panel).
  Shaded regions (hardly visible) mark a combination of sampling and cosmic variance errors for the GR cases.}
\label{fig:matter_Xins}
\end{figure*}
Studying the 1-point density field statistics was useful to obtain general insights into the differences between GR and nDGP
at various scales and density thresholds. However, in the observational context, the 2-point statistics used commonly to study
clustering in cosmological and extragalactic context should be much more illuminative. Fig.~\ref{fig:variance_PT} already hints
at nDGP's effects on the matter field variance. There we can observe that the second moment is enhanced by an approximately 
constant factor on all scales. The amplitude of this increase takes $\sim4\%$ ($\sim20\%$) for $\Omega_{rc}=0.0124$ ($\Omega_{rc}=0.438$) models.
This is of course readily characterised by increased effective $\sigma_8\equiv\sqrt{\sigma^2}(8\hmpc)$ values of the nDGP simulations,
which are higher by $2\%$ and $10\%$ for both nDGP models respectively. The variance of a smoothed density field is a position space
counterpart of a matter density power spectrum $P(k)=\langle \de_k\de_k^*\rangle$, which is a 2-point variance statistic for Fourier modes.
For completeness we show the power spectra alongside their fractional deviations in Fig.~\ref{fig:matter_pk}. The picture fostered here is
consistent with the behaviour exhibited by real-space variance, $\sigma^2$. The degree of the deviation from a GR fiducial value, as well as
the similar scale independence of this enhancement agree well with what we have observed earlier. However, thanks to the decomposition of 
the density field into an orthogonal basis of density fluctuation modes, we can easier depict characteristic features of the Vainshtein screening
mechanism. Namely, at small scales, where the density field variance is dominated by the interiors of virialised haloes, we can start
to observe the suppression of the nDGP enhanced clustering due to effective screening of the fifth-force. This was studied in much greater
detail by other authors (see \eg \cite{MOG_comp}), but we can confirm that our simulations, up to their rather limited mass resolution, abide 
to those findings (see also \cite{Falck2015}).
An important observation here consist of noting that as we have predicted, the simple 2-point statistics is not capable of fully capturing and 
describing the otherwise complicated departures of the nDGP cosmic density field from the fiducial GR case, as was hinted by the PDFs shown in
Fig.~\ref{fig:pdf_compare}. This indicates a need to look at higher-order moments in order to get further insights into the gravitational
evolution of clustering in MG theories with the Vainshtein screening.

\subsubsection{Higher moments and hierarchical cumulants}
\label{subsec:dm:higher_moments}
\begin{figure*}
 \includegraphics[angle=0,width=0.48\textwidth]{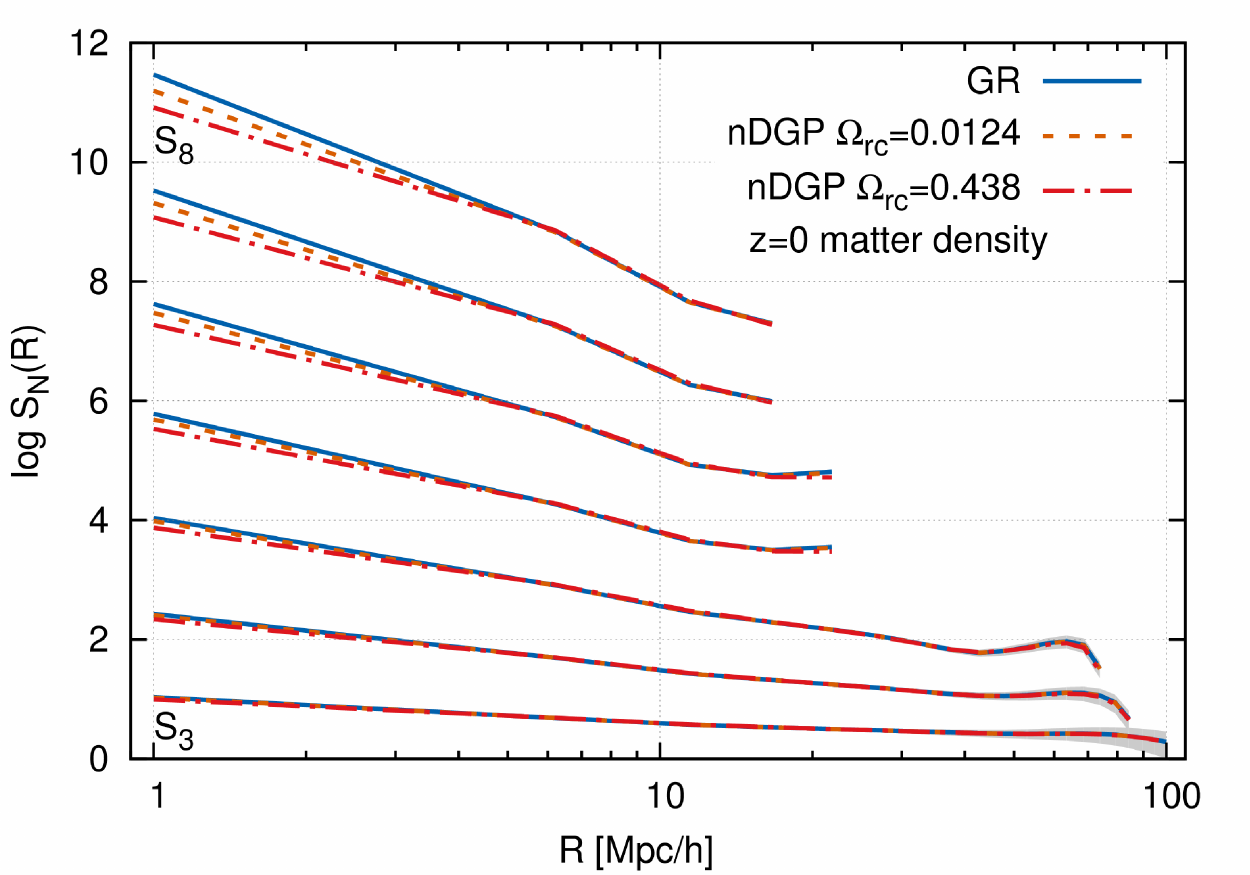}
 \includegraphics[angle=0,width=0.48\textwidth]{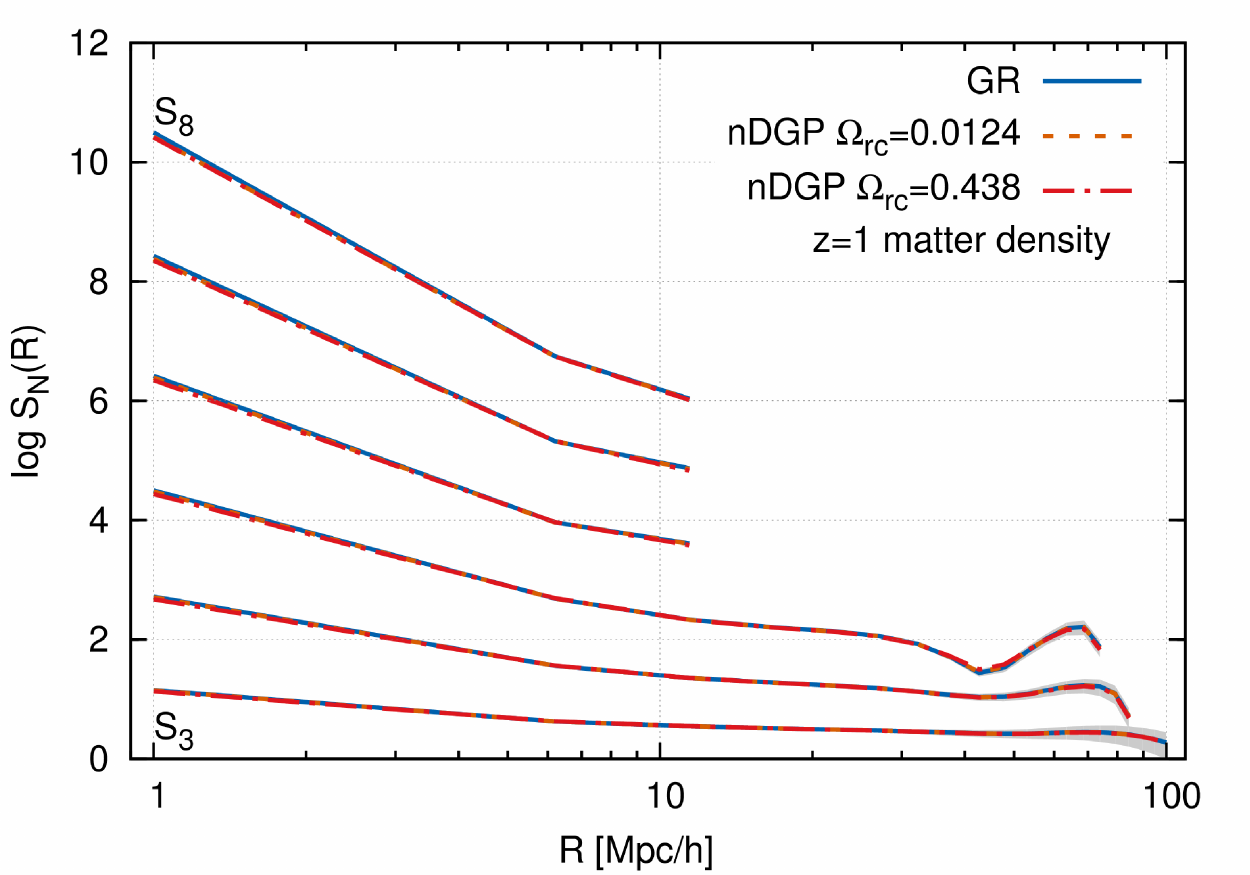}
 \caption{Hierarchical clustering amplitudes $S_n(R)$ computed for smoothed dark matter density fields at $z=0$ (left panel) and $z=1$ (right panel).
 Shaded regions (hardly visible) mark a combination of sampling and cosmic variance errors for the GR cases.}
\label{fig:matter_Sns}
\end{figure*}

In this section we look into the behaviour of higher-order moments of the density fields on a range of cosmic scales and epochs. We first take a look
at the averaged correlation functions themselves before focusing on the reduced cumulants. In Fig.~\ref{fig:matter_Xins} we plot the beautiful 
correlation hierarchy formed by all the central moments from $\overline{\xi}_2$ up to $\overline{\xi}_9$ at $z=0$ (left panel) and $z=1$ (right panel). 
For all our models the consecutive higher-order correlation functions exhibit tiered behaviour with the same monotonic scale-dependence. 
Here, the picture broadly agrees with what we have observed already for the variance alone. Namely, the main effect of the nDGP 
fifth-force is reflected in the $\overline{\xi}_n$'s amplitudes, which are increased by a nearly constant factor over the range of the scales.
We can also comment on a clear emergence of the effects induced by the finite-volume of our simulations. In agreement with well
establish theory\cite{Colombi1994}, the higher-order moments are more severely affected by the limited size of our simulations, which
is indicated by the fact that the moments from $\overline{\xi}_6$ and above could only be reliably estimated up to scales of $\sim 10-20\hmpc$.

The situation gets much more interesting when we contemplate the reduced cumulants, $S_n$'s, which are shown for the same two redshifts
in Fig.~\ref{fig:matter_Sns}. We can observe here the sensitivity of these statistics to the non-linear gravitational evolution,
as was established by many other authors \cite{Juszkiewicz1993,Lokas_kurt,grav_inst_pt,grav_inst_pt2,White1999}. 
In contrast to the connected moments alone, the nDGP-induced deviations of the matter hierarchical 
amplitudes are characterised by a strong scale-dependence. Here, at $z=0$, the deviations take significant values at small scales $R\simlt10\hmpc$, while
going back to the fiducial GR predictions at larger separations. This behaviour is nearly not present at $z=1$, indicating that
the features we observe here in $S_n$'s are induced by non-linear gravitational dynamics, which at late times 
induce the change of the power spectrum shape at small scales, where the Vainsthein mechanism is ta play 
(see \eg{} the high-$k$ tail in the lower panel of Fig.~\ref{fig:matter_pk}).
The ratios for reduced cumulants as described in Eqns.(\ref{eqn:hier_amplitudes}) and (\ref{eqn:pt_gamma}), 
are sensitive by this shape change induced by modified structure formation. Hence we can confirm 
the potential of the density field hierarchical amplitudes as potential probes of modified gravitational dynamics, a finding already
emphasised in clustering studies for a different class of MG theories \cite{Hellwing2010,Hellwing2013}.
\begin{figure*}
 \includegraphics[angle=-90,width=0.98\textwidth]{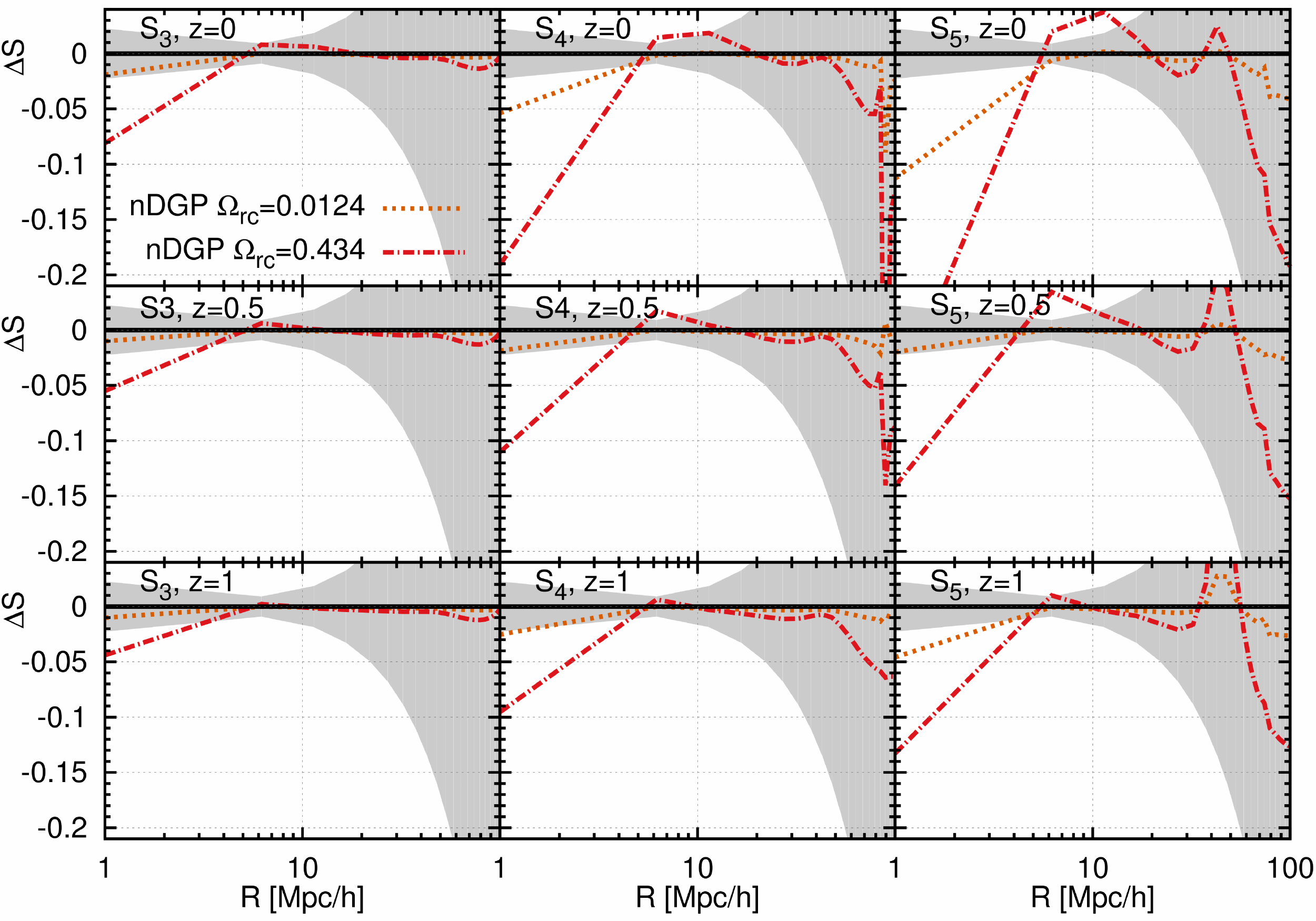}
 \caption{Fractional differences from the GR taken across three epochs $z=0, 0.5, 1$ (rows of panels from top to bottom)
 of the first three matter density reduced cumulants $S_3, S_4$ and $S_5$ (columns of panels from left to right).
 Shaded regions illustrate the total error budget (cosmic variance + shot noise) on the ratios.}
\label{fig:matter_s3s4s5}
\end{figure*}
\begin{figure*}
 \includegraphics[angle=0,width=0.99\textwidth]{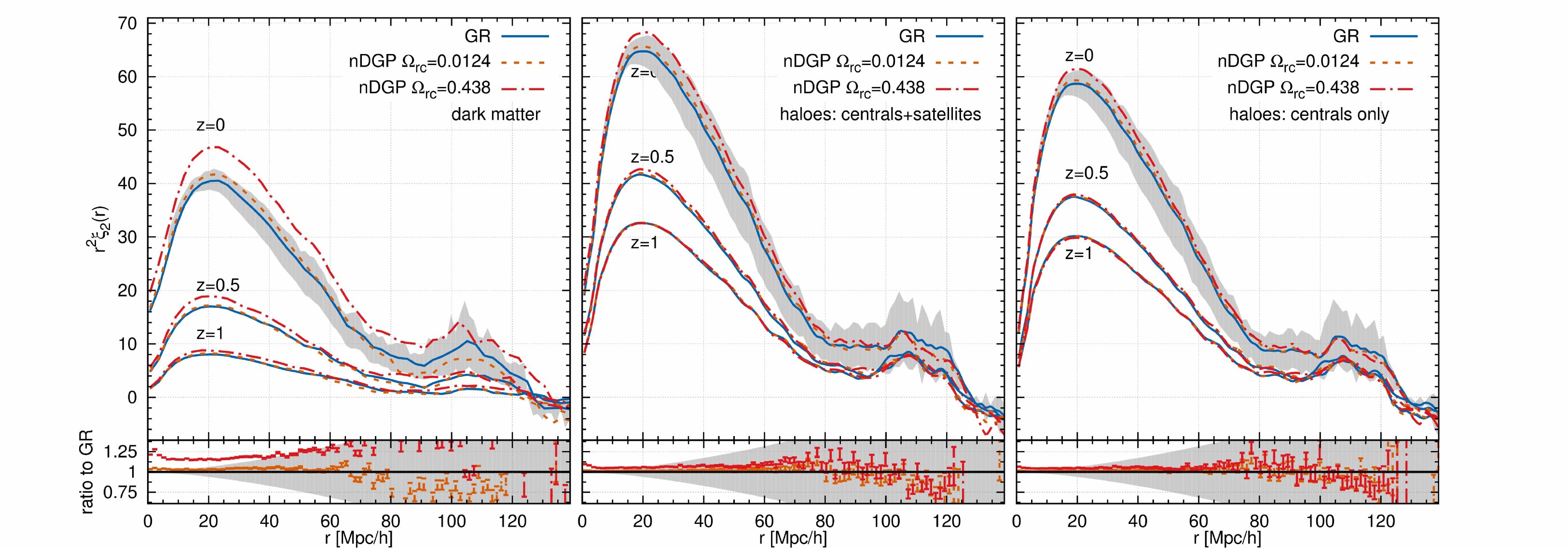}
 \caption{Two-point correlation functions computed as a pair separation function R. {\it Left panel:} Two-point statistics
 for subsampled dark matter particles three our models: GR (solid blue), nDGP $\Omega_{rc}=0.0124$ (dashed orange) and
 nDGP $\Omega_{rc}=0.438$ (dashed-dotted red). The shaded regions mark illustrative cosmic variance error for the GR case.
 Three distinctive epochs are shown: $z=0,0.5$ and $1$. Consecutive values of $r^2\xi_2(r)$ were scaled down to allow better
 presentation.
 {\it Middle panel:} Same as the left panel but for all DM haloes (centrals and satellites) identified in our simulations.
 {\it Right panel:} Same as the middle panel but only for main haloes.
 In all three cases the lower subpanels illustrate the fractional departure from the fiducial GR case ($\xi_2/\xi_2^{GR}-1$)
 at $z=0$. Here the error bars mark the Poison sampling errors for the pair number counts in each $r+\Delta r$ separation bin.
 In contrast the shaded region (shown only for $z=0$ for clarity) reflect the cosmic variance error contribution.
 }
\label{fig:mat_halo_2PCRF}
\end{figure*}

We now want to obtain a more quantitative measures of the general $S_n$'s behaviour observed above. Since our $N>5$ moments are severely 
affected by finite-volume effects at even intermediate scales, in the following we will limit our analysis to only first three 
reduced cumulants: skewness ($S_3$), kurtosis ($S_4$) and $S_5$. We are interested in the fractional differences ($\Delta S$) of the nDGP hierarchical
amplitudes taken as always w.r.t. the fiducial GR case assessed for a range of scales and epochs. We plot the results for the three 
cumulants taken at three redshifts: $z=0,0.5$ and $1$ in the nine panels of Fig.~\ref{fig:matter_s3s4s5}. As already observed in the previous
plots, the non-GR signatures decrease with redshift and increase with the reduced cumulant order. For all the cases except the reduced 
skewness, we can observe some erratic behaviour at large scales. This is driven by the relatively limited volume of our simulation,
and we can attribute this behaviour to noise, as indicated clearly by the cosmic variance shaded areas engulfing all the lines
at $R\simgt 10\hmpc$. However, at large scales there could be potentially interesting features appearing around the BAO wiggle,
which is a common feature predicted to exist at a fraction of the acoustic horizon scale in the higher order moments (see the details 
in \citep{SkewBAO}). Our current limited simulations prevents us from charting this territory, and we will leave the studies of the BAO-related
scales for the future.

Focusing on small scales, where cosmic variance is sub-dominant, we note that there is a strong signal at a level of 
significance $>2-3\sigma$ present for both our nDGP models in $S_4$ and $S_5$. The predicted deviations of the large cross-over scale
model of $\Omega_{rc}=0.0124$ for $S_3$ are too small to be significant in the presence of our sampling noise. However, for the case
of the stronger nDGP model its signature is prominent enough to constitute a $>2\sigma$ strong signal for all redshifts even in the skewness alone.

\subsection{Haloes}
\label{subsec:haloes}
It is always very insightful to trace and study the correlation hierarchy induced by gravitational clustering on the matter density field.
In the previous section we have seen that the fifth-force induced by the scalar-field propagation in the nDGP cosmology affect 
the higher-order moments of the matter density field only at relatively small scales. This was consistent with the behaviour we could
trace in the full 1D pdf's of the density field. However, in reality the matter density field is not accessible to us directly
from observations. This is why we will now analyse the clustering hierarchy exhibited by dark matter halo populations as found
in our simulations. As described in Sec.\S\ref{sec:simulation} we split our DM haloes into various populations, based on their spatial
averaged abundances. In this way, we can study the clustering of different density tracers, and in principle such halo populations
can be (after making some simplifying assumptions) related to various populations of observed luminous galaxies. We shall not attempt
to model here any more complicated effects that are related to a given galaxy survey selection, nor any geometry effects.
This would require construction of a dedicated survey-specific mock galaxy catalogue and our current simulations are too limited in
volume and resolution to allow for such a procedure. However, in the future, once more advanced and bigger MG simulations become
available, it will be worthwhile to apply our analysis to survey-dedicated mocks. 

\subsubsection{2-point statistics}
\label{subsec:haloes:2pt_stat}
\begin{figure}
 \includegraphics[angle=0,width=0.46\textwidth]{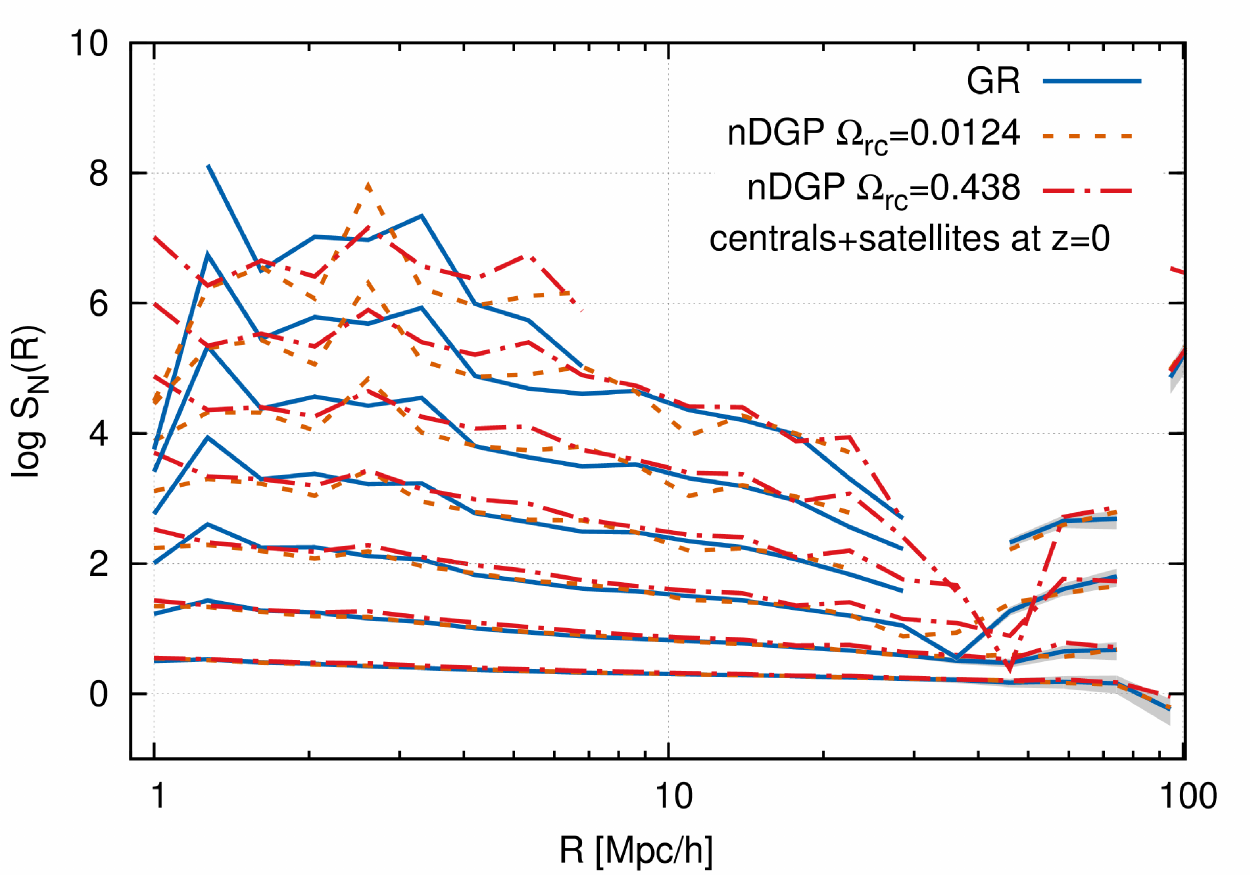}\\
 \includegraphics[angle=0,width=0.46\textwidth]{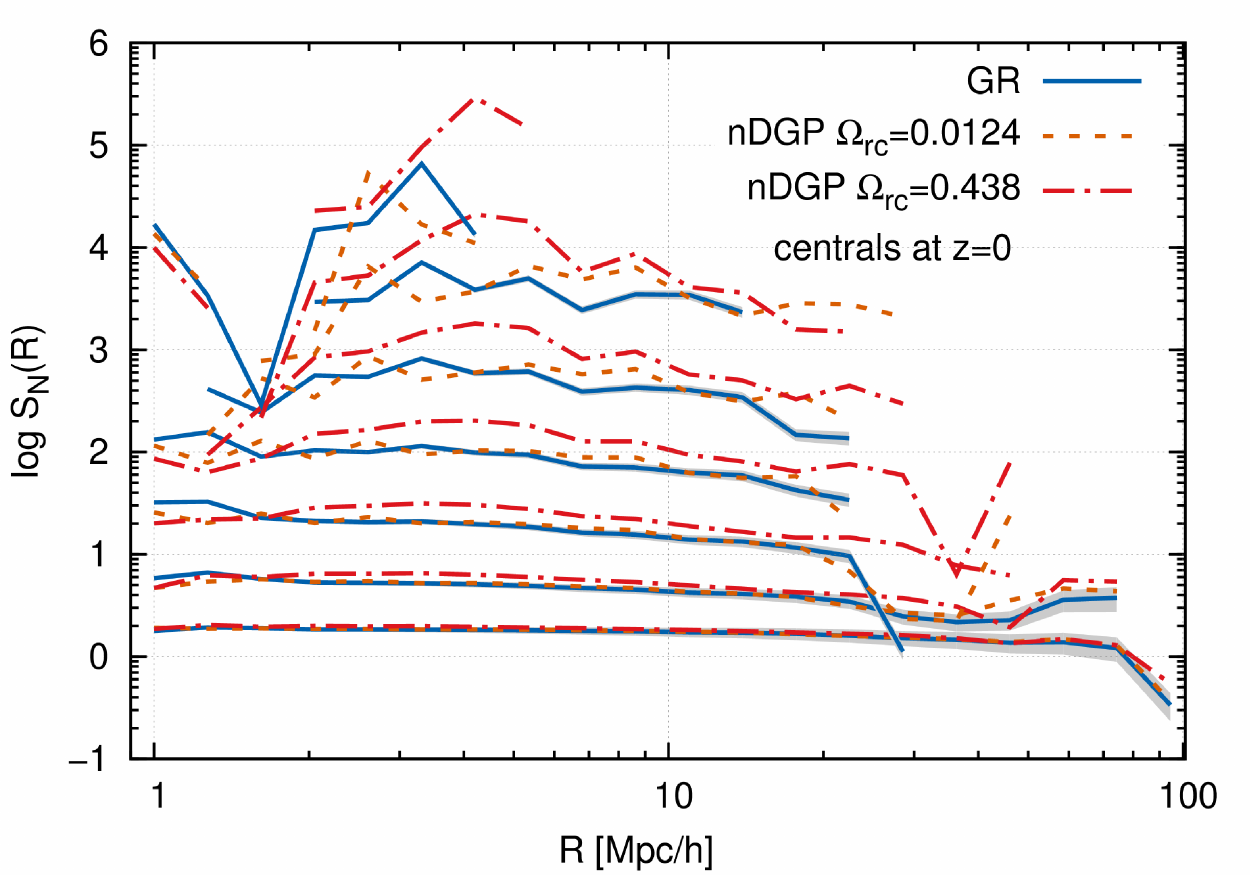}
 \caption{The hierarchical amplitudes tiered  from the reduced skewness, $S_3$, (the most bottom group of lines in each panel) 
 to $S_9$ (the top group of lines in each panel)  calculated for the DM halo overdensity field at $z=0$. In the top panel 
 we show the data for all haloes found in our simulations (centrals+sattellites),  while in the bottom panel we use 
 main haloes only (centrals).
 }
\label{fig:haloes_Sns}
\end{figure}
We begin with taking a look at the 2-point clustering statistics expressed in terms of the configuration space 2-point correlation
functions $\xi(r)$. In Fig.~\ref{fig:mat_halo_2PCRF} we plot in three panels, the scaled $r^2\xi(R)$ for the dark matter sub-sample
with $\overline{n}=1.8\times10^{-3}h^3/$Mpc$^3$ (left panel), the central+satellites halo sample with $\overline{n}=1.4\times10^{-3}h^3/$Mpc$^3$ at $z=0$
(middle panel) and the centrals only sample with $\overline{n}=1.3\times10^{-3}h^3/$Mpc$^3$ (right panel). We consider the three epochs
$z=0,0.5$ and $1$, with the higher redshifts lines downscaled for brevity. As usual we show the relative ratios to the fiducial GR case
in the bottom panels, here displayed only for the $z=0$ case, where the deviations from GR are the strongest.
Since now we have individual pair-number counts for each model in each separation bin $r+\Delta r$, we decide
to show the Poisson error bars reflecting given number-counts separately from the cosmic variance error (shaded region). For the matter $\xi_2$
we observe that the nDGP deviations appear on all scales probed, however due to large cosmic variance plaguing our simulations,
the $1\sigma$ significance is limited to pair-separation scales of $r\simlt 15(40)\hmpc$ for nDGPa (nNDGb) models. 
The situation for  both our most-abundant halo populations is in stark contrast. Firstly, we need to note that the relative differences are much smaller.
While for the DM 2PCF the model with the smallest cross-over scale attained a difference of a $25\%$ magnitude, for haloes the maximum departure
from the GR 2PCF amplitude is around $5\%$, except for the smallest separations $r\sim1-3\hmpc$ where it reaches $10\%$ strength.
We have also check the 2PCF amplitudes for our more diluted halo samples. There, the error due to sparse-sampling shot noise is much more
severe rendering all the differences from the GR at small scales to be statistically insignificant. 

This was a very important exercise. The study of the $\xi_2(r)$ amplitudes has revealed that in the case of nDGP gravity the strong signal
present in the matter density field clustering gets strongly suppressed and nearly diminishes when one look at the clustering of DM haloes.
There still might be a possibility to identify and extract a more robust MG signal in the 2PCF alone. For example one might try to use a specific
combinations of $\xi_2$ amplitudes taken at different scales and for different galaxy/halo samples such as the clustering ratios advocated
in \cite{Arnalte-Mur2017}. However, a detailed study of this kind would require simulations with more realisations, so the cosmic variance contribution would 
be minimised. Thus we postpone it for the future, when such data sets will become available.

\subsubsection{Higher moments and hierarchical cumulants}
\label{subsec:healoes:higher_moments}
\begin{figure*}
 \includegraphics[angle=-90,width=0.96\textwidth]{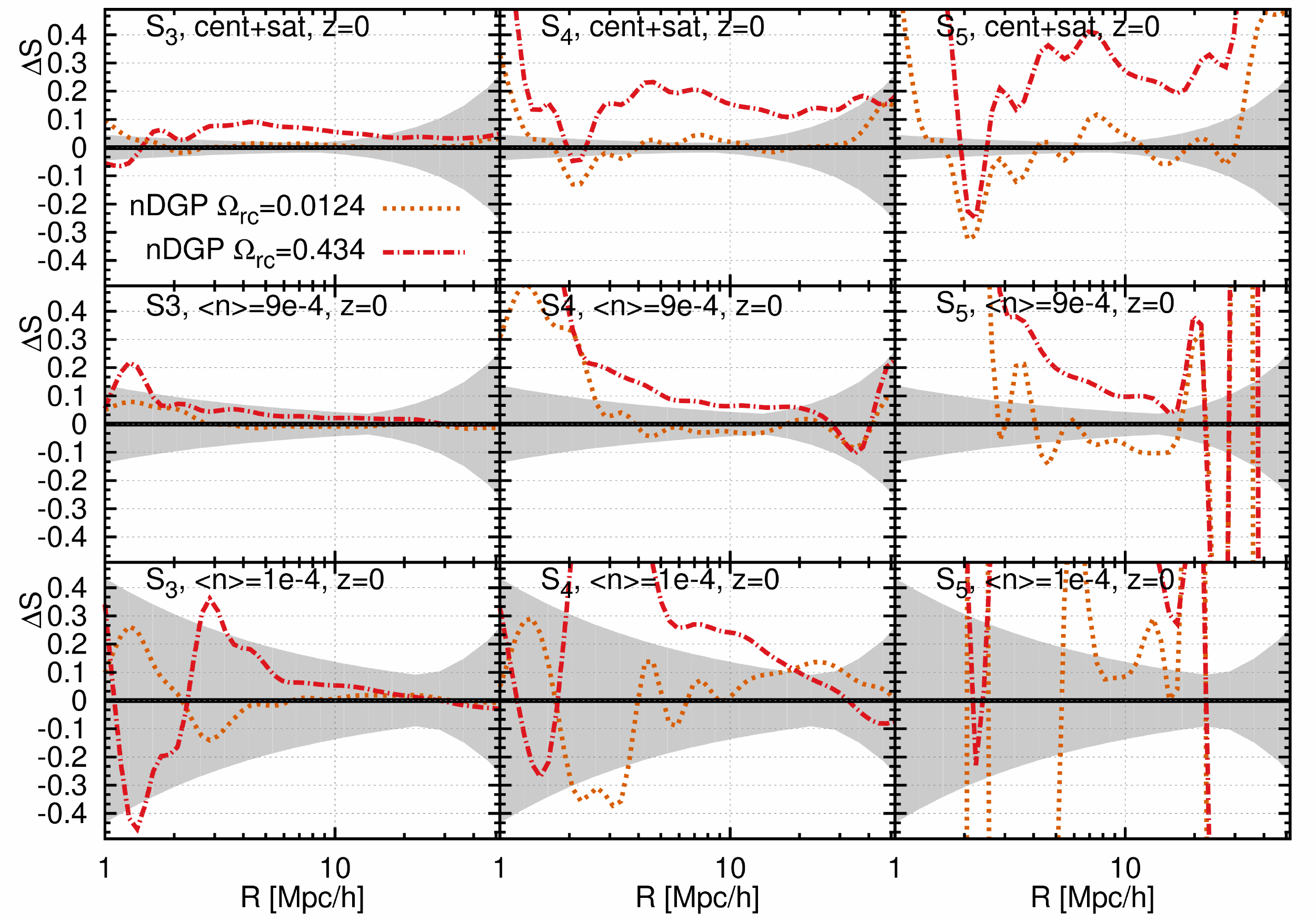}
 \caption{Fractional differences from the GR taken at $z=0$ for three halo samples:
 $\av{n}=1.4\times10^{-3}h^3/$Mpc$^3$(centrals+satellites), $\av{n}=9\times10^{-4}h^3/$Mpc$^3$
 and $\av{n}=1\times10^{-4}h^3/$Mpc$^3$ (rows of panels from top to bottom)
 of the first three reduced cumulants $S_3, S_4$ and $S_5$ (columns of panels from left to right).
 Shaded regions illustrate the total error budget (cosmic variance + shot noise for each halo sample) on the ratios.}
\label{fig:haloes_S345}
\end{figure*}

As we have mentioned a couple of times already the two-point statistics are not enough to tell the full story. As in the case of the DM clustering,
we can expect that MG effects on the halo clustering statistics might be better visible in the higher-order moments. The general picture for the case
of plain central averaged moments $\overline{\xi}_n$ is very similar to the one we have observed for the matter clustering moments shown in
Fig.~\ref{fig:matter_Xins}, thus we will skip this and jump straight away to the reduced cumulants $S_n$'s. We plot them at $z=0$ for two 
halo samples (centrals only and centrals+satellites) in Fig.~\ref{fig:haloes_Sns}. First, we would like to focus on the impact made by removing 
the satellites from the halo sample. This effect is clearly visible in the lower panel for $R\simlt 3\hmpc$, where especially the higher-order
cumulants starting from $S_6$ experience a dramatic suppression of amplitude at those scales. This effect is caused by halo exclusion. By removing
the subhaloes from our sample we are unable to probe the 1-halo regime any more. Thus at scales that are comparable with halo sizes we are left
with only two limiting cases, one or zero halo count in a given region of space. This is a well known effect \cite{Bardeen1986,Baldauf2013}.
However, an important observation
can be made here. By studying the scale and the magnitude of the halo exclusion effect we have found that they are in quantitative agreement
for all three models we study here. This is consistent with the very similar satellites fractions we have found for all the three simulations
(see Sec.\S\ref{sec:simulation}).

The second important feature fostered by the results shown in Fig.~\ref{fig:haloes_Sns} is that the differences between models are much larger
and also extend to larger scales than what we have observed for the matter field. The additional contrast here is the fact that now the differences
from GR can take both negative and positive sign, while they only were negative in the case of DM. As we will see later, especially for 
the $\Omega_{rc}=0.438$ case, for most of the scales considered the reduced cumulants take actually larger amplitudes than in the GR case. A behaviour which
is exactly opposite to the one we have witnessed for the matter density hierarchical amplitudes. Since the reduced cumulants are scaled by 
the averaged variance this effect ought to be driven mostly by higher-order biasing of haloes.

To obtain a more quantitative insight into the nDGP effects we will focus on the first three cumulants ($S_3-S_5$) estimated at $z=0$ for which
we find the strongest differences. We show the relative deviations from GR of these hierarchical amplitudes for three halo samples in Fig.\ref{fig:haloes_S345}.
Since the ratios get too noisy to allow for a more robust analysis we take an average 3-bin centred value and use cubic splines to smooth
the data.
The general trend that the non-GR signature gets stronger as we consider higher moments is also confirmed. Interestingly, we observe that
the relative differences here can take much larger values than in the case of DM cumulants and they are carried up to much larger scales.
While for the matter density all cumulants started to converge on the GR values for $R\simgt 10\hmpc$, here for all our halo samples
the differences can still be as large as $10-40\%$ up to $R\simlt 50\hmpc$.

For the case of the reduced skewness the difference is only significant for the stronger nDGPb model and only for the richest
one of our halo samples (centrals+satellites), where the magnitude of non-GR signal is varying from $5\%$ to $10\%$ at the scales we consider.
Despite the fact that $\Delta S_3$ is relatively small, there is a significant departure from the GR mean, 
reaching for $4\leq R(\hmpc)^{-1}\leq 9$ a $3\sigma$ statistical strength. 
By considering two main halo samples only ($\av{n}=9\times 10^{-4}h^3/$Mpc$^{3}$ and 
$\av{n}=1\times 10^{-4}h^3/$Mpc$^{3}$) we observe, that 
the relative differences from the GR case are larger for the sparser of the two samples. We need to conclude however that the errors connected 
with the shot-noise and sparse-sampling are so severe that the non-GR signature is enveloped by the noise for the case 
the two central halo samples. We can expect that going to higher spatial abundances (hence lower-mass haloes) would
improve the situation. However, a new class of higher-resolution simulations are needed to perform such a study. It is also clear that for
all three halo samples the $\Omega_{rc}=0.0124$ signature is too weak to be significant. It emerges above the sampling-noise error envelop
only at the smallest probed scale ($R=1\hmpc$) for centrals+satellites sample, alas with a statistical signal less than $2\sigma$.

Now we want to focus on the reduced kurtosis, since it appears that this cumulant can be potentially the most promising observable to look
for a significant non-GR signature. Here all halo samples contain a significant nDGP signal at different scales .
For the large cross-over scale of nDGPa model we find that at $R\simlt 3\hmpc$ the signal reaches a few $\sigma$ significance.
For the stronger nDGPb model, the signal is significant and very prominent for all scales when we consider the centrals+satellites
sample and up to $R\simlt20\hmpc$ for centrals with $\av{n}=9\times 10^{-4}h^3/$Mpc$^{3}$. For the lowest density sample the departure from 
the GR case is so strong that it is still significant in the regime $2\simlt R/(\hmpc)\simlt20$. This is a very promising result, as 
such a number density of tracers can be already attained by for example the Luminous Red Galaxy (LRG) sample in the SDSS/BOSS survey \cite{SDSS_LRGs}.
Thus, our finding open an exciting possibility to use the reduced kurtosis of LRG clustering as a discriminatory test for non-GR theories
involving the Vainshtein screening. For the large cross-over scale model we observe that it departures from the GR case in a significant way only
at small scales. For the main halo sample at $R\leq 3\hmpc$ the signal becomes marginally ($2-3\sigma$) strong.

Finally, we move to $S_5$, which fosters, as predicted, the largest relative deviations from the GR case. Alas, at larger separations ($R\simgt20\hmpc$)
and for sparser samples the ratio of nDGP to GR becomes very erratic and starts to fluctuate around zero. This reflects both the fact that
our simulations are characterised by relatively low volume coverage and the fact that higher-order moments are very vulnerable to sparse sampling noise.
Nonetheless, for our richest sample the non-GR signature reaches $\sim40\%$($\sim10\%$) for nDGPb(nDGPa) at $R\sim8-10\hmpc$.
The corresponding statistical significances are $8\sigma$  and $3\sigma$ respectively. For the centrals only with $\av{n}=9\times 10^{-4}h^3/$Mpc$^{3}$
we find that the strong nDGP model experiences a significant departure from the GR for $R\simlt20\hmpc$, where at the smallest scales
($R\simlt5\hmpc$) the statistical significance for the $S_5$ reaches up to $6-7\sigma$ from the GR mean. Here, for the same sample, 
the mild nDGP model at larger separations fluctuate around the GR case, but again at the small scales $<3\hmpc$ the deviation become strong, reaching 
magnitude of $5\sigma$. For both models and at all scales the ratio of $S_5$ experiences a large scatter with the $\av{n}=1\times 10^{-4}h^3/$Mpc$^{3}$
sample. Here, the signal could in principle be strong enough for a prospective detection, but our simulations are not good enough to allow us
to make that claim.

We can conclude that our analysis indicates that the reduced kurtosis, $S_4$, appears to offer a most beneficial combination of the non-GR signature
amplitude and noise properties to maximise the signal strength. While $S_5$ fosters usually stronger relative differences, it appears to be 
really noisy for our halo samples as well. However in principle, one could imagine that with a data sample covering a much larger volume, the
impact of the finite-volume effect would be suppressed, increasing the quality of the signal in $S_5$. We also see that for most of the cases,
if one considers only a very mild MG model, one needs to go to really small scales ($R\simlt5\hmpc$) to look for a prominent signal. 
At such small scales, deep into the non-linear regime, the modelling of galaxy clustering becomes very complicated due to various 
degeneracies connected with baryonic effects and non-linear galaxy bias \cite{skewness_barions,mog_gadget,vanDaalen2014,Hellwing2016}.

\section{Discussion and Conclusions}
\label{sec:conclusions}
In this paper we have studied the gravitational instability mechanism of the GR and two nDGP models as our chosen test-case examples
of MG with the Vainshtein screening mechanism. To follow the gravitational evolution into the non-linear regime 
we have conducted and analysed a series of N-body simulations implementing WMAP9 cosmogony and the non-linear dynamics 
of GR and nDGP with $\Omega_{rc}=0.0124$ and $0.438$. Specifically, we have focused on using both low and high-order clustering statistics
to study the matter density field and halo clustering across the cosmic epochs (from $z=1$ to $z=0$) and scales (from $1$ to $100\hmpc$).
For the haloes we have considered a few different populations characterised by a different spatial abundances 
(from $\av{n}=1.9\times10^{-3}h^3/$Mpc$^{3}$ to $1\times10^{-4}h^3/$Mpc$^{3}$). Thus, in the spirit of the abundance matching, we have made
a first attempt towards modelling of observables that can be easier associated with the different observational samples of many galaxy surveys.
Here we provide a list of our findings constructed to emphasise the most important ones:
\begin{itemize}
 \item For all considered models and density field statistics we have found that the maximal relative difference from the GR case is always attained at
 $z=0$ and is generally a monotonic function of time. 
 \item The investigation of one dimensional $p(\de+1)$ functions has revealed that the pdfs of all models take maximums at the same averaged density values,
 but are characterised by different shapes of low and high density tails. The most prominent was the increased MG pdf width 
 (reflected in measured later larger variance). We also noticed a significantly larger asymmetry (especially for the smallest smoothing scale $R=0.5\hmpc$). 
 In all cases and for all epochs the 1-point density statistics indicated that the nDGP matter density fields are characterised by enhancements at both 
 lower and higher density tails. 
 \item Both the matter density field variance and the power spectrum fostered consistent picture. Here, the non-GR models are characterised by a nearly
 constant and scale-independent enhancement factors of $4$ and $20\%$ for our mild and strong nDGP models respectively. 
 \item We found a remarkable agreement of the PT-based estimators for the variance and the skewness with our N-body results. The absolute amplitudes
 are in percent-level agreement down to $10(15)\hmpc$ for the variance (skewness). At the same time the relative deviations form GR retained the accuracy
 even down to smaller scales of $\sim5\hmpc$.
 \item The high-order moments of the matter field, $\overline{\xi}_2-\overline{\xi}_9$ were only mildly affected by the MG dynamics. However,
 the corresponding hierarchical amplitudes $S_3-S_8$ are marked with significant deviations from the GR case at small scales $R\simlt 10\hmpc$
 for the $z=0$. At higher redshifts the differences are typically sizeable smaller. The main feature is that nDGP models are characterised
 by smaller amplitudes $S_n$ w.r.t. the GR fiducial case at small scales, while they converge to the GR values at larger separations (\ie $R\simgt 10\hmpc$).
 This reflect the fact that at larger scales the matter power spectrum ($P(k)$) and variance ($\sigma^2(R)$) are enhanced by a constant factor, but
 the nDGP gravity does not change their shape.
 \item In the 2PCF the relative deviations from the GR were observed to be much stronger for the DM case than in the halo sample. With the deviations 
 in the matter clustering stretched out to even large scales $\sim 10\hmpc$. This was not found to be the case for the halo 2PCF, where the significant 
 differences were contained only to the smallest pair separations $R\simlt10\hmpc$.
 \item The non-GR signal was much stronger in higher-order halo clustering statistics. Reduced moments are characterised by departures from GR-fiducial
 values up to much larger scales ($R\simgt 50\hmpc$) than in the case of DM clustering. A side remark would be that the halo exclusion effect was 
 marked prominently in higher $S_n$'s for the centrals-only sample.
 \item Qualitatively we have found that the skewness offers a good chance constrain only strong nDGP models (like our nDGPb) and only for 
 our halo sample with the highest spatial density. On the other hand, our study has pinpointed the reduced kurtosis as an excellent candidate for
 a very promising cosmological MG probe. Here the nDGPb signal is prominent on all nearly scales and for all halo samples. 
 The model with large cross-over scale nDGPa is also harbouring strong signal, alas contained to only small scales of $R\simlt3\hmpc$. In general,
 we were able to identify the non-GR signal with a good statistical significance, varying from 3 to $8\sigma$ for a given model.
 \item Lastly, we have observed even stronger relative signal in $S_5$, however this was accompanied by a much larger sampling noise effectively reducing
 the statistical significance of it. We advertise however, that with a better and larger-volume sample it might become feasible to extract
 MG signal from $S_5$ with a large statistical significance.
\end{itemize}

In the literature we find abundant studies pointing toward a very similar picture as fostered by our analysis. In general, we see a trend 
in dark matter clustering properties in models exhibiting a non-negligible fifth-force on intergalactic scales, and this 
is shared among models despite a specific screening mechanism they employ. Namely, we observe that the fifth-force acting on DM
in MG theories enhances the gravitational instability process. This is reflected in a more efficient transfer of matter from underdense
to ovedense regions, which results in cosmically depressed regions, like cosmic voids, to be characterised 
by lower densities \cite[see \eg][]{Hellwing2009,halosvoids_fr,Clampitt2013}.
Due to mass conservation the average density remains the same as in the GR case, hence the first affected clustering characteristic
is of the second-order, thus the variance of the density field. We saw that both in the configuration space variance as well as in the 
power spectrum. The gravitational instability also naturally leads to more complicated shape deviations of the density field 
distribution function. This is reflected in growing amplitude of the asymmetry rank -- the  skewness,
or tails squashing as measured by the kurtosis. To understand why the fifth force dynamics, while enhancing the density variance w.r.t.
to the fiducial GR case, actually leads to lower skewness and kurtosis (and higher-order cumulants as well), we need to recall that 
there is an intrinsic asymmetry imprinted in the density field described by the density contrast $\de$. While on the positive side of 
the distribution the density can grow to arbitrarily large values, reaching presently for example orders of $\sim10^{6}$ in the very centres
of cluster sized DM haloes, on the negative size there is a fundamental barrier of $\de=-1$, which reflect the limiting case of a zero density 
in an absolutely empty region of space. Thus in MG scenarios, and in the nDGP in particular, more empty voids lead to reduced asymmetry
of the PDF thanks to the enhanced $\de<0$ tail of the distribution.  One could argue that in general the more emptier voids are compensated
by increased density inside haloes and filaments. But since the latter are occupying much smaller fraction of the Universe, the overall 
effects of voids prevails in the higher-order cumulants.

The DM density field is not accessible to observations, but one is able to measure the distribution of the total matter by using weak lensing
techniques. It has been already proven that weak lensing statistics such as the lensing convergence $\kappa$ and its power spectrum $C_{\kappa\kappa}$
are useful cosmological probes.
Our results for the matter density moments indicate that the non-GR signature we found should also be present 
in weak lensing statistics \cite{Schmidt2008,Beynon2010}.
However, it would be very hard to assess the feasibility of the weak lensing statistics as probe of nDGP-kind of MG. This is because,
as our studies have found, most of the MG induced effects are confined to relatively small scales $R\simlt 10\hmpc$. There, the baroynic effects
related to energetic processes of galaxy formation are a source of a major influence on the total matter distribution 
\cite{skewness_barions,Hellwing2016}. A separate dedicated study, involving self-consistent hydrodynamical simulations with galaxy 
formation physics are needed to test the robustness and usability using the distortions of matter clustering as a MG probe.

The second part of the analysis presented in this work yields however a much more optimistic picture. Our studies of the higher-order 
halo clustering statistics have shown that there is a very appealing opportunity to use the reduced cumulants as cosmological probes of non-GR
models and simultaneously consistency checks for GR. Here, especially the reduced kurtosis appears as the most promising statistics.
Since, we have not attempted any realistic galaxy modelling beyond some simple abundance matching approach, we should caution that
there are still a number of potential systematics effects related to both galaxy formation and survey selection that can potentially
weaken the MG signal we have found. However, the fact that the strong signal was present in all three halo samples we considered,
is an optimistic indicator that the MG signature persists over a range of halo masses. Thus, we can expect that it should be also present
in galaxy samples constructed using various selection criteria. 

Another source of potential worry consists of a notion that the clustering statistics that are measured from spectroscopic 
galaxy surveys concerns the position of galaxies in the redshift space. In the present study we have only focused on position space
clustering neglecting the effects of halo peculiar velocities. In MG theories we can expect in general that the fifth-force is affecting
the comic velocity field much stronger than the density alone \cite{LiHell2013,Hellwing2014PhRvL}, this may lead to a signal degeneracy at scales where
the both clustering and halo velocities are strongly affected. Thus, redshift space distortions contrary to classical approach, where they
are a source of rich cosmological information, in the case of the reduced cumulants are a source of another systematic.
Withal, it has been shown that in the case of the hierarchical amplitudes, the redshift space distortion appears to affect the variance
and the higher order moments to a similar degree, and since these two quantities appears in the denominator and numerator of $S_n$ formula
respectively, the overall redshift space effect is suppressed \cite{Hivon1995}. Thus, one can hope that the signal we have identified in this work
would persists also in the redshfts space cumulants of galaxy clustering.

Therefore, we strongly advocate here the urgent need for a good quality galaxy mock catalogues for MG theories that would allow to test
for the robustness of MG signal in galaxy $S_n$'s. Once such data sets will become available, we will be given the opportunity to 
exploit already rich galaxy clustering data from the surveys like 
SDSS/BOSS\cite{Ross2006}, 2dF\cite{Baugh2004,Croton2004_2dF} or VIPERES\cite{Cappi2015} and even more powerful data that will
come from future grand-scale surveys like Euclid and DESI, to perform an independent and robust tests of GR and non-GR models on
intergalactic scales.

\section*{Acknowledgements}
The authors are very grateful to Baojiu Li for inspiring discussions and for providing the {\tt ECOSMOG} code 
that was used to run the simulations in this paper.
WAH and KK are supported from European Research Council (grant number 646702 ``CosTesGrav'').
WAH also acknowledges support from the Polish National Science Center under contract \#UMO-2012/07/D/ST9/02785.
KK is also supported by the UK Science and Technologies Facilities Council 
grants ST/N000668/1. BB was supported by University of Portsmouth.
GBZ is supported by NSFC Grant No. 11673025, and by a Royal Society-Newton Advanced Fellowship.
The simulation used in this work were carried on the Sciama HPC cluster on the University of Portsmouth.
This work used the DiRAC Data Centric  system at Durham University,
operated by the Institute for  Computational Cosmology on behalf of 
the STFC DiRAC HPC Facility (\url{www.dirac.ac.uk}). This equipment 
was funded by BIS National E-infrastructure capital grant ST/K00042X/1, 
STFC capital grant ST/H008519/1, and STFC DiRAC Operations grant 
ST/K003267/1 and Durham  University. DiRAC is part of the National 
E-Infrastructure. This research was carried out with the support of the 
HPC Infrastructure for Grand Challenges of Science and Engineering 
Project, co-financed by the European Regional Development Fund under 
the Innovative Economy Operational Programme.
\renewcommand{\bibname}{References}
\bibliographystyle{h-physrev}
\bibliography{nDGP_hierar}

\end{document}